\documentclass[%
 reprint, prl
 amsmath,amssymb,
 aps, longbibliography
]{revtex4-1}

\usepackage{graphicx}
\usepackage{dcolumn}
\usepackage{bm}
\usepackage{tipa}
\usepackage{upgreek}

\usepackage{amsmath}
\usepackage{bm}
\usepackage{xcolor}
\usepackage{soul}
\usepackage[normalem]{ulem}
\usepackage{siunitx}
\usepackage[version=4]{mhchem} 
\usepackage[capitalize]{cleveref}
\usepackage{float}

\usepackage{xcolor, soul}
\definecolor{darkpurple}{RGB}{102,0,153}
\sethlcolor{yellow}

\begin{document}

\title{Coupled self-charging and self-propulsion of evaporating sessile droplets}


\author{Nitish Singh$^{1}$}
\author{Aaron D. Ratschow$^{2}$}
\author{ Syed N. R. Kazmi$^{3}$}
\author{Dan Daniel$^{1}$}
    \email{dan-daniel@oist.jp}

\affiliation{$^{1}$Droplet and Soft Matter Unit, Okinawa Institute of Science and Technology Graduate University, Onna, Okinawa 904-0495, Japan}
\affiliation{$^{2}$Max Planck Institute for Polymer Research, 55128 Mainz, Germany}
\affiliation{$^{3}$Nanofabrication Core Lab, King Abdullah University of Science and Technology (KAUST), Thuwal 23955-6900, Saudi Arabia}

\begin{abstract}
Evaporating sessile water droplets universally deposit surface charges while simultaneously becoming self-charged---a phenomenon that remains under-appreciated despite its potentially important consequences. Here, we investigate one such consequence: droplet self-propulsion driven by self-charging. This self-propulsion arises spontaneously in the absence of external stimuli and produces self-avoiding trajectories that can displace droplets by many times their own diameter. Our results have important practical implications for processes requiring spatial precision, such as inkjet printing. The coupling between self-charging and self-propulsion produces mosaic surface charge patterns characterized by alternating polarities, which we explain with a physical model based on properties of the electric double layer.  
\end{abstract}

\maketitle
 
\section{Introduction}
Few physical phenomena are as common, or as deceptively simple, as a droplet drying on a surface \cite{wilson2023evaporation}. This everyday phenomenon, known as sessile droplet evaporation, is in fact a rich, multi-scale problem that governs processes ranging from coffee-ring and blood pattern formations \cite{deegan1997capillary, brutin2011pattern, Popov2005} to spray cooling \cite{kim2007spray} and inkjet printing \cite{lohse2022fundamental}. While conventionally viewed as a heat and mass transfer problem \cite{Picknett1977, Hu2002, erbil2012evaporation, gelderblom2022evaporation}, sessile droplet evaporation can be profoundly influenced by electrostatics---a largely unexplored dimension that is the focus of this work.
 
Previously, we and others showed that evaporating water droplets universally deposit surface charges at the receding contact line \cite{He.2019, Knorr.2024, Bipolar_singh_2025}, due to charge transfer from the electric double layer (EDL) \cite{Stetten.2019, Li.2022, Ratschow.2024, Ratschow.2025}. As a result, even an initially uncharged droplet becomes self-charged. Here, we demonstrate a previously unrecognized consequence of this self-charging process: spontaneous self-propulsion directed by the surface charge landscape. 

For directed self-propulsion to occur, the axial symmetry of the droplet retraction must be broken to produce a net lateral force. In previous systems, symmetry breaking is imposed externally---through vibrations \cite{brunet2007vibration_external, daniel2002rectified_external}, geometric confinement \cite{prakash2008surface_geometry}, wettability or thermal gradients \cite{brzoska1993motions, daniel2001fast_wettability_gradient}, printed surface charge patterns \cite{Sun.2019}, or the addition of surfactants \cite{toyota2009self_chemical_reaction, hanczyc2007fatty_chemical_rxn}. By contrast, here, symmetry breaking occurs spontaneously for \textit{pure} water droplets in the absence of external stimuli, mediated by contact-angle hysteresis \cite{Mettu2011, cha2025pinning}. During evaporation, random pinning sites produce non-axisymmetric contact-line retraction dynamics and, consequently, a non-axisymmetric surface charge landscape, generating a net electrostatic force that drives propulsion. Since surface charging is commonly observed on dielectric substrates \cite{WANG201934, Ratschow.2025}, the mechanism is broadly relevant. 

The droplet trajectory is guided by its charging history, with the surface charge landscape deposited during prior evaporation directing propulsion that in turn shapes the subsequent charge patterns. This strong coupling between surface charge deposition, droplet self-charging, and droplet self-propulsion gives rise to self-avoiding trajectories and mosaic surface charge patterns characterized by alternating polarities. Notably, droplets can travel many times their own diameter, with implications for the spatial precision of technologies such as inkjet printing.

We propose a physical model that accurately reproduces the surface charge patterns from the evaporation and propulsion dynamics, based on two key properties of the EDL, namely its charge density $\sigma_{\text{EDL}}$ and charge transfer coefficient $\alpha$ \cite{Stetten.2019,Ratschow.2025}. By balancing the electrostatic force generated by the surface charge landscape against the pinning force arising from contact-angle hysteresis, our model predicts an explicit criterion for the onset of propulsion, explaining why droplet self-propulsion is observed on some surfaces (e.g., fluoropolymer) but not others (e.g., acrylic).

Our findings have important implications, including for evaporation-driven electrical energy generation \cite{xue2017water,du2025water}, deposition of colloids and molecules \cite{aizenberg2000patterned,thakur2013directed,Zhou.2025}, and potentially acceleration of chemical reactions on charged surfaces \cite{chen2022water, li2026chemical}.       

\section{Electrostatic origin of self-propulsion}

\begin{figure*}[!htb]
\centering
\includegraphics[scale=1]{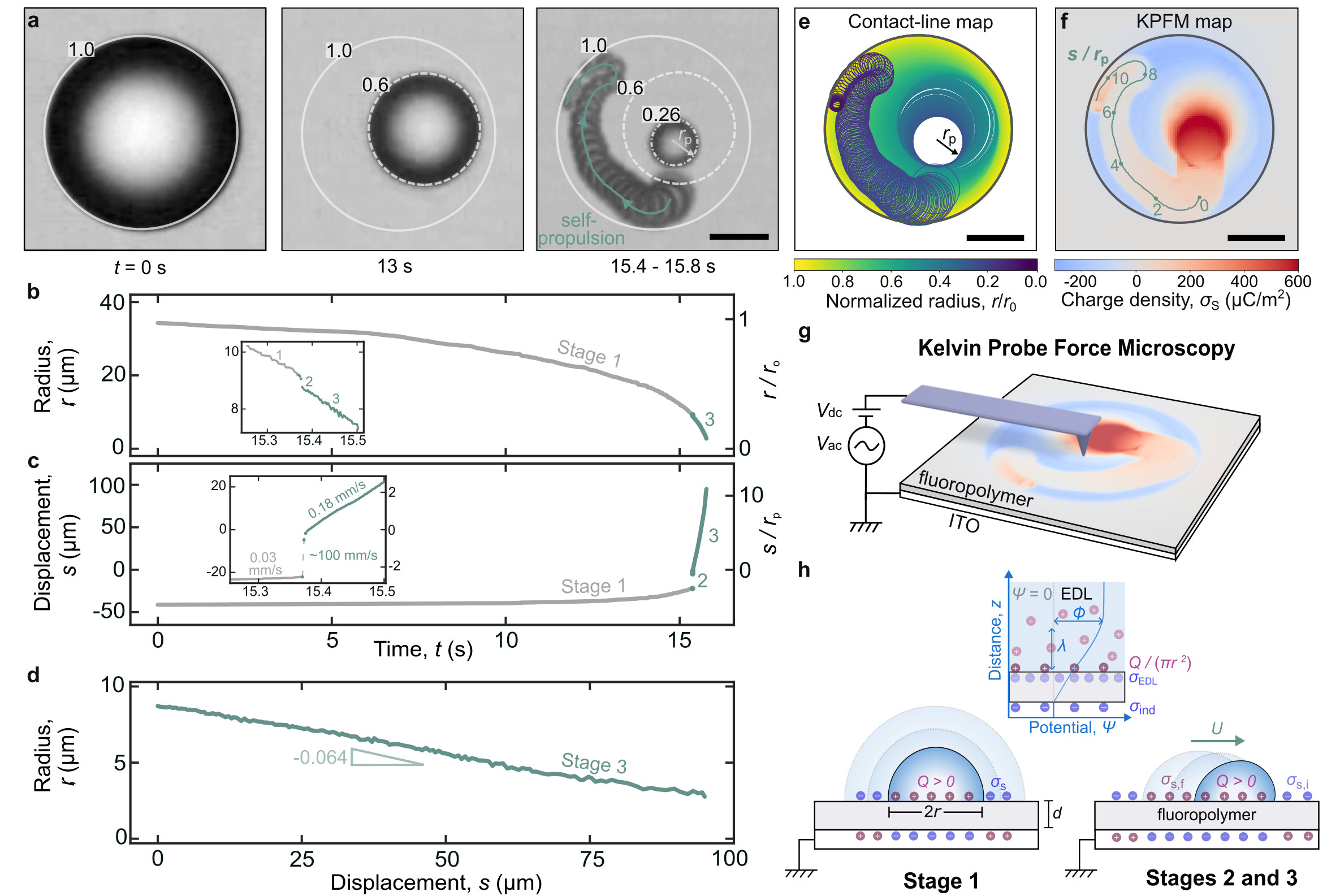}
\caption{\label{fig:phenomenon} \textbf{Electrostatic origin of self-propulsion}. (a) Timelapse images of droplet evaporation and self-propulsion, with labels indicating the normalized radius $r/r_{0}$. (b) Evolution of droplet radius $r$ and (c) displacement $s$ with time; insets show detail near the onset of self-propulsion. (d) $r$ vs. $s$ during self-propulsion. The correspondence between the maps of (e) contact-line positions (parameterized by $r/r_{0}$) and (f) surface charge reveals the electrostatic origin of self-propulsion. Scale bars in a, e, and f: \SI{20}{\micro\meter}. (g) Surface charges are mapped using KPFM. (h) Schematic of the electric double layer (EDL): charge transfer from the EDL at the receding contact line leads to droplet self-charging, which in turn drives self-propulsion.}
\end{figure*}

Deionized water microdroplets were sprayed onto a fluoropolymer film of thickness $d = \SI{220}{\nano\meter}$, supported on a grounded indium tin oxide (ITO) substrate [Fig.~\ref{fig:surface}]. The water droplets were polydisperse, with initial footprint radius $r_0 = 29\pm7\,$\SI{}{\micro\meter}, and carried minimal charges $|Q_{0}| \lesssim \SI{20}{\femto\coulomb}$ [Fig.~\ref{fig:distribution}]. 
  
Fig.~\ref{fig:phenomenon}(a) shows a representative timelapse of such an evaporating droplet at an ambient temperature of \SI{22}{\celsius} and 50\% relative humidity. For most of its lifetime, evaporation of the sessile droplet is relatively uneventful (Stage 1, $t$ = 0--15.3 s): during this time, the droplet footprint radius $r$ decreases gradually from \SI{34}{\micro\meter} to \SI{9}{\micro\meter} [Fig.~\ref{fig:phenomenon}(b)], while its center shifts slightly, covering a distance of $s$ = \SI{40}{\micro\meter} at speeds $U <$ \SI{0.03}{\milli\meter\per\second} [Fig.~\ref{fig:phenomenon}(c)]. See Supporting Video S1.

Once the droplet reaches the propulsion radius $r_{p} = \SI{9}{\micro\meter}$ ($r/r_0 = 0.26$), it undergoes an abrupt onset of propulsion, covering $s = \SI{16}{\micro\meter}$ within \SI{2.5}{\milli\second} and reaching a peak speed of $U \sim \SI{100}{\milli\meter\per\second}$ (Stage 2, $t = \SI{15.38}{\second}$) [inset in Fig.~\ref{fig:phenomenon}(c); Fig.~\ref{fig:fast}]. This sudden burst is followed by sustained self-propulsion at a comparatively steady $U = \SI{0.18}{\milli\meter\per\second}$ (Stage 3, $t$ = 15.4--15.8 s), over which the droplet moves by $s = \SI{100}{\micro\meter}$, equivalent to about ten times its radius $r_{p}$. Interestingly, the self-propelling droplet avoids its own previous path [third panel of Fig.~\ref{fig:phenomenon}(a); later in Fig.~\ref{fig:phenomenon}(e)]; this self-avoiding trajectory is also bounded by the initial footprint of the droplet [circle corresponding to $r/r_{0} = 1$ in Fig.~\ref{fig:phenomenon}(a); black outer circles in Fig.~\ref{fig:phenomenon}(e),(f); Supporting Video S2]. During stage 3, $r$ decreases linearly with traveled distance $s$, with slope $|\mathrm{d}r/\mathrm{d}s| = 0.064$ [Fig.~\ref{fig:phenomenon}(d)]. 


The same evaporation and self-propulsion dynamics are well captured by the contact-line map in Fig.~\ref{fig:phenomenon}(e), showing both the contact-line positions and normalized radius $r/r_{0}$ at different times. During stage 1, successive contact-line positions do not cross one another. By contrast, as the droplet self-propels during stages 2 and 3, the receding contact line at a later time crosses the advancing contact line from an earlier time.

\begin{figure*}[!htb]
\centering
\includegraphics[scale=1]{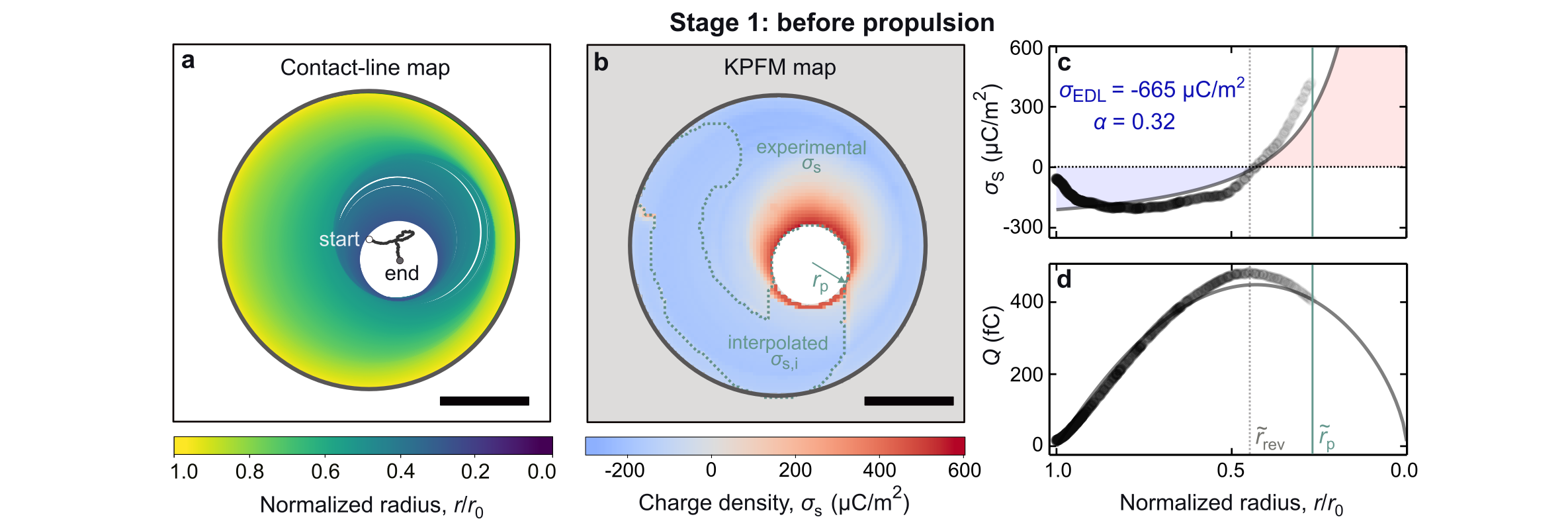}
\caption{\label{fig:before} \textbf{Surface charge deposition before propulsion}. (a) Contact-line map and (b) KPFM map before the onset of propulsion, i.e., for $r > r_{p}$. Scale bars: \SI{20}{\micro\meter}. (c) $\sigma_{\text{s}}$ and (d) $Q$ as functions of $r/r_{\text{0}}$. Lines represent theoretical predictions from Eqs.~\eqref{eq:sigma_anal} and \eqref{eq:Q_anal}, while symbols denote experimental values from (b). Standard deviations are smaller than the symbol size.}
\end{figure*}

This contact-line map can be directly compared with the surface charge map $\sigma_{\text{s}}(x, y)$ [Fig.~\ref{fig:phenomenon}(f)], spatially resolved using Kelvin probe force microscopy (KPFM) immediately after droplet evaporation [Fig.~\ref{fig:phenomenon}(g)]. Importantly, the receding contact-line positions coincide with contours of constant $\sigma_{\text{s}}$, a correspondence that points to the electrostatic origin of droplet self-propulsion.  

The deposited surface charge pattern, which we attribute to charge transfer from the EDL [Fig.~\ref{fig:phenomenon}(h)], exhibits several notable features. First, it is bipolar: the droplet initially deposits negative surface charge, followed by positive charge at later stages. Second, the deposition is non-monotonic during self-propulsion [path indicated by line in Fig.~\ref{fig:phenomenon}(f)], with $\sigma_{\text{s}}$ oscillating in both magnitude and polarity, resulting in mosaic charge patterns. This contrasts with conventional slide electrification, where a droplet deposits unipolar charges that decay monotonically with distance \cite{Slide_electic_Stetten2019, bista2023high, ratschow2025liquid}. The origin of these features will be explained in the remainder of the manuscript. 

\section{Stage 1: before propulsion \label{sec:stage1}}

We start by describing the surface charge deposition process during stage 1, i.e., in the absence of self-propulsion \cite{Bipolar_singh_2025}. The surface charge density deposited at the receding contact line is given by
\begin{equation}
	    \sigma_\mathrm{s} = \alpha \left( \sigma_\mathrm{EDL} + \frac{Q}{\pi r^{2}}\right).
 \label{eq:sigma_s}
\end{equation}
This deposited charge therefore has two components: one from the EDL, $\sigma_{\text{EDL}}$, and another from the droplet's net charge, $Q/(\pi r^2)$ [Fig.~\ref{fig:phenomenon}(h)]. The charge transfer coefficient $0 \leq \alpha \leq 1$ quantifies the fraction of charge transferred. Implicit in Eq.~\eqref{eq:sigma_s} is the assumption that the droplet charge $Q$ is localized at the droplet base rather than distributed over its spherical-cap surface. This simplification is justified because the parallel-plate capacitance $\varepsilon_{0} \varepsilon_{s} (\pi r^{2}/d)$ is much larger than the spherical-cap capacitance $\sim \pi\varepsilon_{0} r$ for $r \gg d$; here $\varepsilon_0$ is the vacuum permittivity and $\varepsilon_s = 2$ the relative permittivity of the fluoropolymer film.

As the droplet deposits charge at density $\sigma_{\text{s}}$, it must simultaneously acquire $\mathrm{d}Q = 2 \pi r \sigma_\mathrm{s} \mathrm{d}r$; this conservation of charge then yields an ordinary differential equation that describes the self-charging process
\begin{equation}
    \frac{\mathrm{d}Q}{\mathrm{d}r}=2\alpha \left( \pi r \sigma_\mathrm{EDL} + \frac{Q}{r}\right), \label{eq:ODE_Q}
\end{equation}
which, for an initially uncharged droplet, has the non-dimensionalized solutions
\begin{equation}
	\tilde{\sigma}_{\text{s}} (\tilde{r}) = \frac{\alpha}{1-\alpha} \left( 1 - \alpha \, \tilde{r}^{2\alpha-2} \right)
	\label{eq:sigma_anal}
\end{equation}
and
\begin{equation}
	\tilde{Q} (\tilde{r}) = \frac{\alpha}{1-\alpha} \left( \tilde{r}^2 - \tilde{r}^{2\alpha} \right),
	\label{eq:Q_anal}
\end{equation}
where $\tilde{r} = r/r_0$, $\tilde{\sigma}_{\text{s}} = \sigma_{\text{s}}/\sigma_{\text{EDL}}$, and $\tilde{Q} = Q/(\pi r_0^2 \sigma_{\text{EDL}})$; here and throughout, non-dimensional quantities are denoted by a tilde. As long as contact-line positions do not cross one another, both $\tilde{\sigma}_{\text{s}}$ and $\tilde{Q}$ depend only on $\alpha$ and $\tilde{r}$, independent of retraction dynamics (axisymmetric and non-axisymmetric). 

\begin{figure*}[!htb]
\centering
\includegraphics[scale=1]{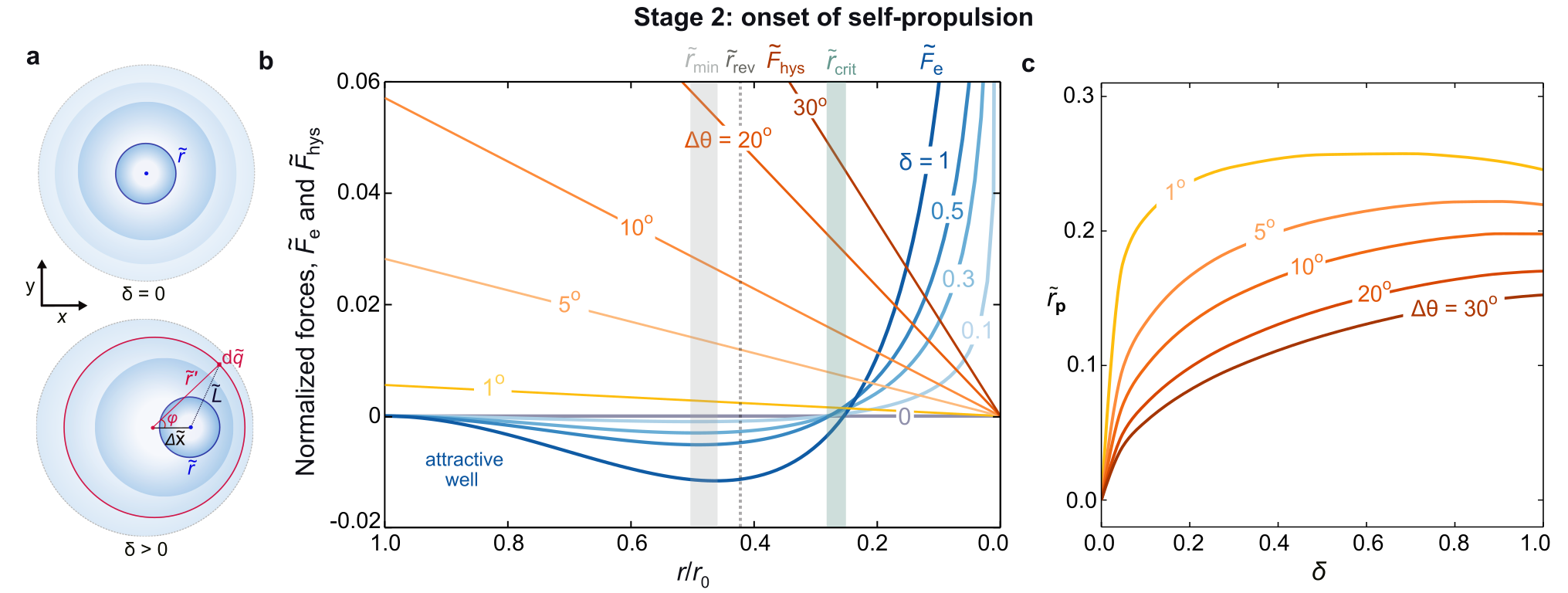}
\caption{\label{fig:onset} \textbf{Onset of propulsion}. (a) Schematic of droplet shift parameterized by $\delta$. (b) Normalized electrostatic force $\tilde{F}_{e}(\tilde{r})$ (blue, Eq.~\eqref{eq:Fe}) for several values of $\delta$, and normalized pinning force $\tilde{F}_{\text{hys}}(\tilde{r})$ (orange, Eq.~\eqref{eq:Fhys}) for several values of $\Delta\theta$. (c) Lines show the theoretical onset radius $\tilde{r}_{\text{p}}(\delta)$ given by the crossing point of $\tilde{F}_{e}$ and $\tilde{F}_{\text{hys}}$.}
\end{figure*}

The analytic solutions make concrete predictions, which can be tested against experimental results. Fig.~\ref{fig:before}(a) and (b) reproduce the contact-line and $\sigma_{\text{s}}$ maps of Fig.~\ref{fig:phenomenon}(e) and (f), respectively, but truncated at $r = r_{p}$. Within the region traversed later by the self-propelled droplet [bounded by dashed lines in Fig.~\ref{fig:before}(b)], we show the interpolated values $\sigma_{\text{s,i}}(r)$ obtained by radially averaging experimental $\sigma_{\text{s}}(r)$ outside this region [Fig.~\ref{fig:before}(c)]. The droplet charge $Q(r)$ can be obtained by numerically integrating $\sigma_{\text{s}}(x, y)$ in Fig.~\ref{fig:before}(b), including interpolated values, outside the droplet footprint   
\begin{equation}
Q(r) = -\iint_{\text{outside}} \sigma_{\text{s}}(x, y) \, \mathrm{d}A. \label{eq:charge_integral}
\end{equation}

At the start of the evaporation process ($\tilde{r} = 1$), Eq.~\eqref{eq:sigma_anal} predicts $\sigma_{\text{s}} = \alpha\, \sigma_{\text{EDL}}$, i.e., the droplet deposits charge of the same polarity as $\sigma_{\text{EDL}}$, which in this case is negative [Fig.~\ref{fig:before}(c)]. At the same time, the droplet acquires positive $Q$ [Fig.~\ref{fig:before}(d)]. However, upon reaching the reversal radius
\begin{equation}
\begin{split} \label{eq_r_rev}
\tilde{r}_{\text{rev}} = (1/\alpha)^{1/(2\alpha - 2)},
\end{split}
\end{equation}
$\sigma_{\text{s}}$ reverses polarity, giving rise to bipolar charging. This is also the point where $Q$ reaches its maximum value.  From the measured $\tilde{r}_{\text{rev}} = 0.43$, Eq.~\eqref{eq_r_rev} gives $\alpha = 0.32$; further fitting the measured $Q(r)$ to Eq.~\eqref{eq:Q_anal} then yields $\sigma_{\text{EDL}} = \SI{-665}{\micro\coulomb\per\meter\squared}$. With these two fitted parameters, Eqs.~\eqref{eq:sigma_anal} and \eqref{eq:Q_anal} reproduce the measured $\sigma_{\text{s}}(r)$ and $Q(r)$ with good quantitative agreement over the range $r < r_{p}$ [Fig.~\ref{fig:before}(c),(d)]. We previously showed that the bipolar surface charging is a general feature of evaporating sessile droplets, allowing us to extract $\alpha$ and $\sigma_{\text{EDL}}$ for different materials, such as plastics and glass \cite{Bipolar_singh_2025}. 

The magnitude and sign of $\sigma_{\text{EDL}}$ reflect the surface potential $\phi$ of the EDL \cite{Vogel.2022}. In the Debye--H\"uckel regime (small $\phi$), the two are simply related by $\sigma_{\text{EDL}} = \varepsilon_{0} \varepsilon_{\text{L}} \phi/\lambda$, where $\lambda \approx$ 200 nm is the Debye length and $\varepsilon_{\text{L}} = 80$ is the relative permittivity of the liquid (water). We also assume that $\alpha$ is constant, independent of contact-line velocity. This holds because contact-line motion during stage 1 is slow with $\dot{r} \sim \SI{10}{\micro\metre\per\second}$, which corresponds to a P\'eclet number $\text{Pe} = \dot{r}\lambda/D_{\text{ion}} < 10^{-2}$, assuming an ionic diffusivity $D_{\text{ion}} \approx 10^{-9}\,\text{m}^{2}\,\text{s}^{-1}$. Since $\text{Pe} \ll 1$, the diffusive equilibration timescale within the EDL, $\lambda^{2}/D_{\text{ion}}$, is much shorter than the advective timescale $\lambda/\dot{r}$, and charge transfer is a quasi-static process.


Finally, a closer inspection shows that the charge pattern deposited during stage 1 is non-axisymmetric [Fig.~\ref{fig:before}(b)] due to a shift of the droplet center [line in Fig.~\ref{fig:before}(a)]. This spatial asymmetry produces a net electrostatic force on the droplet that drives self-propulsion, which we explain in the next section.

\section{Stage 2: onset of propulsion}

We introduce a parameter $0 \leq \delta \leq 1$ that sets the maximum droplet shift during stage 1, which is assumed to be rectilinear in the positive $x$ direction. As the droplet radius $\tilde{r}$ shrinks, the droplet center shifts to $\tilde{x} = (1-\tilde{r})\delta$, where the coordinate $\tilde{x}$ is likewise normalized by the initial footprint radius $r_{0}$. Hence, $\delta = 0$ corresponds to fully axisymmetric retraction, while $\delta = 1$ corresponds to the maximally asymmetric case.

Since every past position of the receding contact line is a circle, the charge landscape seen by a droplet of current radius $\tilde{r}$ is a nested family of rings of radius $\tilde{r}' > \tilde{r}$. Each ring carries a constant charge density $\tilde{\sigma}_{\text{s}}(\tilde{r}')$ but has non-uniform width $(1-\delta\cos\varphi)\,\mathrm{d}\tilde{r}'$, where $\varphi$ is the polar angle about the ring center. The center of each ring is located at $(1-\tilde{r}')\delta$; hence, there is an offset of $\Delta\tilde{x} = \delta(\tilde{r}' - \tilde{r})$ between the respective centers of each ring and the droplet [Fig.~\ref{fig:onset}(a)]. Because the rings are not concentric with the droplet, each exerts a net electrostatic force, which we evaluate below.

A charge element $\mathrm{d}\tilde{q}(\tilde{r}', \varphi) = \tilde{\sigma}_{\text{s}}(\tilde{r}')\,\tilde{r}'\,\mathrm{d}\varphi\,(1-\delta\cos\varphi)\,\mathrm{d}\tilde{r}'$ on one of these rings contributes
\begin{equation}
\begin{split}
	\mathrm{d}\tilde{F}_{e}(\tilde{r}, \tilde{r}', \varphi) = \frac{\tilde{Q} \, \mathrm{d}\tilde{q}}{\tilde{L}^2} \left[ \frac{2}{1 + \varepsilon_{s}} - \frac{1}{\varepsilon_{s}} \right] \left[\frac{\Delta\tilde{x} - \tilde{r}'\cos\varphi}{\tilde{L}} \right],
\end{split}
\label{eq:dFe}
\end{equation}
where
\begin{equation}
\begin{split}
	\tilde{L}(\tilde{r}, \tilde{r}', \varphi) &= \sqrt{ (\tilde{r}'\cos\varphi - \Delta\tilde{x})^{2} +  \tilde{r}'^{2} \sin^{2}\varphi }
\end{split}
\label{eq:L}
\end{equation}
is the distance between the droplet center and the charge element. Here and throughout, forces are normalized by $(r_{0}^2 \sigma_{\text{EDL}}^2)/(4 \varepsilon_{0})$.
 
The deposited charge density $\tilde{\sigma}_{\text{s}}(\tilde{r}')$ and the droplet charge $\tilde{Q}(\tilde{r})$ follow from Eqs.~\eqref{eq:sigma_anal} and \eqref{eq:Q_anal}. The term $2/(1 + \varepsilon_{s})$ accounts for the surface charge sitting at the air-dielectric interface, while $-1/\varepsilon_{s}$ accounts for the induced image charge in the grounded ITO layer, $\sigma_{\text{ind}} = -\sigma_{\text{s}}$ [schematic in Fig.~\ref{fig:phenomenon}(h)]. Here, $[\Delta\tilde{x} - \tilde{r}'\cos\varphi]/\tilde{L}$ is the direction cosine that projects the force onto the $x$ direction; the $y$-component vanishes by symmetry following integration.

Integrating $\mathrm{d}\tilde{F}_{e}$ around one ring gives the geometric kernel
\begin{equation}
\begin{split}
	\mathcal{G}(\tilde{r}, \tilde{r}') &= \int_{0}^{2\pi} \frac{(1-\delta\cos\varphi)(\Delta\tilde{x} - \tilde{r}'\cos\varphi)}{\tilde{L}^{3}}\,\mathrm{d}\varphi,
\end{split}
\label{eq:G}
\end{equation}
which can be represented in closed form in terms of complete elliptic integrals \cite{Zypman2006, Ciftja2009} [Appendix~\ref{app:G}].
The total electrical force acting on the droplet then follows by summing $\mathcal{G}$ over the entire deposition history
\begin{equation}
\begin{split}
	\tilde{F}_{e}(\tilde{r}) &= \tilde{Q}(\tilde{r})\left[ \frac{2}{1 + \varepsilon_{s}} - \frac{1}{\varepsilon_{s}} \right] \int_{\tilde{r}}^{1} \tilde{\sigma}_{\text{s}}(\tilde{r}')\, \tilde{r}' \, \mathcal{G}(\tilde{r}, \tilde{r}')\, \mathrm{d}\tilde{r}',
\end{split}
\label{eq:Fe}
\end{equation}
which we implement numerically in Fig.~\ref{fig:onset}(b) for $\alpha = 0.32$ and different $\delta$ values (blue lines). $\tilde{F}_{e}$ acts purely in the $x$ direction, with positive values corresponding to a repulsive force and negative values to an attractive force.

$\tilde{F}_{e}(\tilde{r})$ shows several notable features. First, for $\delta = 0$ (perfectly axisymmetric retraction), $\tilde{F}_{e}(\tilde{r}) = 0$ for all $\tilde{r}$, so self-propulsion cannot occur. Second, for $\delta > 0$, $\tilde{F}_{e}$ is initially attractive, as the droplet deposits \textit{negative} $\sigma_{\text{s}}$ and becomes itself positively charged. The attraction is strongest at $\tilde{r}_{\text{min}} = 0.46$--$0.50$, slightly before the polarity reversal at $\tilde{r}_{\text{rev}} = 0.43$. Upon crossing $\tilde{r}_{\text{rev}}$, the droplet deposits \textit{positive} $\sigma_{\text{s}}$, so the attraction weakens and eventually turns repulsive once $\tilde{r}$ falls below $\tilde{r}_{\text{crit}} = 0.25$--$0.28$. Both $\tilde{r}_{\text{min}}$ and $\tilde{r}_{\text{crit}}$ decrease monotonically with increasing $\delta$, but only weakly, by about 10\% over the full range $0 < \delta \leq 1$ [Table~\ref{tab:rcrit}].

For the droplet to self-propel, the electrostatic force has to overcome the pinning force 
\begin{equation}
\begin{split}
	\tilde{F}_{\text{hys}}(\tilde{r}) = \left( \frac{4 \varepsilon_{0} \gamma}{r_{0} \sigma_\mathrm{EDL}^2 } \right)   2 \tilde{r} \,  \Delta \cos \theta,
\end{split}
\label{eq:Fhys}
\end{equation}
arising from contact-angle hysteresis, $\Delta\theta = \theta_{\text{adv}} - \theta_{\text{rec}}$, where $\theta_{\text{adv}}$ and $\theta_{\text{rec}}$ are the advancing and receding contact angles, and $\Delta \cos \theta = \cos \theta_{\text{rec}} - \cos \theta_{\text{adv}}$. Fig.~\ref{fig:onset}(b) plots $\tilde{F}_{\text{hys}}(\tilde{r})$ for different $\Delta\theta$ values (orange lines), assuming a fixed $\theta_{\text{adv}} = 110^{\circ}$, together with $\sigma_{\text{EDL}} = \SI{-665}{\micro\coulomb\per\meter\squared}$, $r_{0} = \SI{34}{\micro\meter}$, and $\gamma = \SI{72}{\milli\newton\per\meter}$, the surface tension of water.

\begin{figure}[!htb]
\centering
\includegraphics[scale=1]{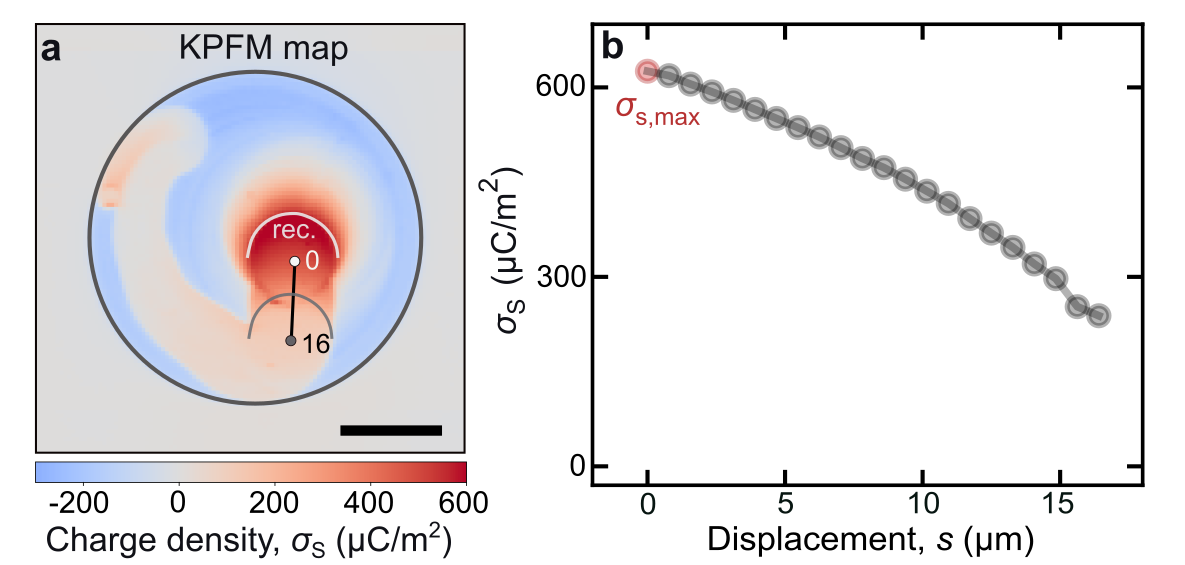}
\caption{\label{fig:sigma_max} \textbf{Charge deposition during stage 2}. (a) KPFM map with the receding contact-line positions at the beginning and end of the fast-motion event indicated. Scale bar: \SI{20}{\micro\meter}. (b) Surface charge density $\sigma_{\text{s}}$ deposited.}
\end{figure}

The onset of propulsion occurs at radius $\tilde{r}_{\text{p}}$ when $\tilde{F}_{e} > \tilde{F}_{\text{hys}}$, i.e., where the $\tilde{F}_{e}$ and $\tilde{F}_{\text{hys}}$ curves cross in Fig.~\ref{fig:onset}(b). The resulting theoretical $\tilde{r}_{\text{p}}(\delta)$ curves for different $\Delta\theta$ values (orange lines) are shown in Fig.~\ref{fig:onset}(c); $\tilde{r}_{\text{p}}$ decreases with increasing $\Delta\theta$ and decreasing $\delta$. Assuming $\Delta\theta = 5^\circ$ and $\delta = 0.5$, we expect propulsion to occur at $\tilde{r}_{\text{p}} = 0.21$. Experimentally, we found that $\tilde{r}_{\text{p}} = 0.27 \pm 0.06$ for 19 different droplets [Table~\ref{tab:rp_exp}], closer to the upper limit of $\tilde{r}_{\text{crit}} = 0.28$ predicted by the theory. The agreement is reasonable given that we assume a rectilinear droplet shift, whereas the true trajectory is curved [line in Fig.~\ref{fig:before}(a)].

The role of $\Delta\theta$ is however more profound than merely setting a threshold to be overcome. Since $\tilde{F}_{e}$ is attractive for $\tilde{r} > \tilde{r}_{\text{crit}}$ and tends to restore symmetric evaporation, asymmetry cannot arise in the absence of contact-angle hysteresis. Paradoxically, then, the same $\Delta\theta$ that must be overcome for propulsion to occur is also what permits the initial symmetry breaking in the first place \cite{Mettu2011, cha2025pinning}.

\begin{figure*}[!htb]
\centering
\includegraphics[scale=1]{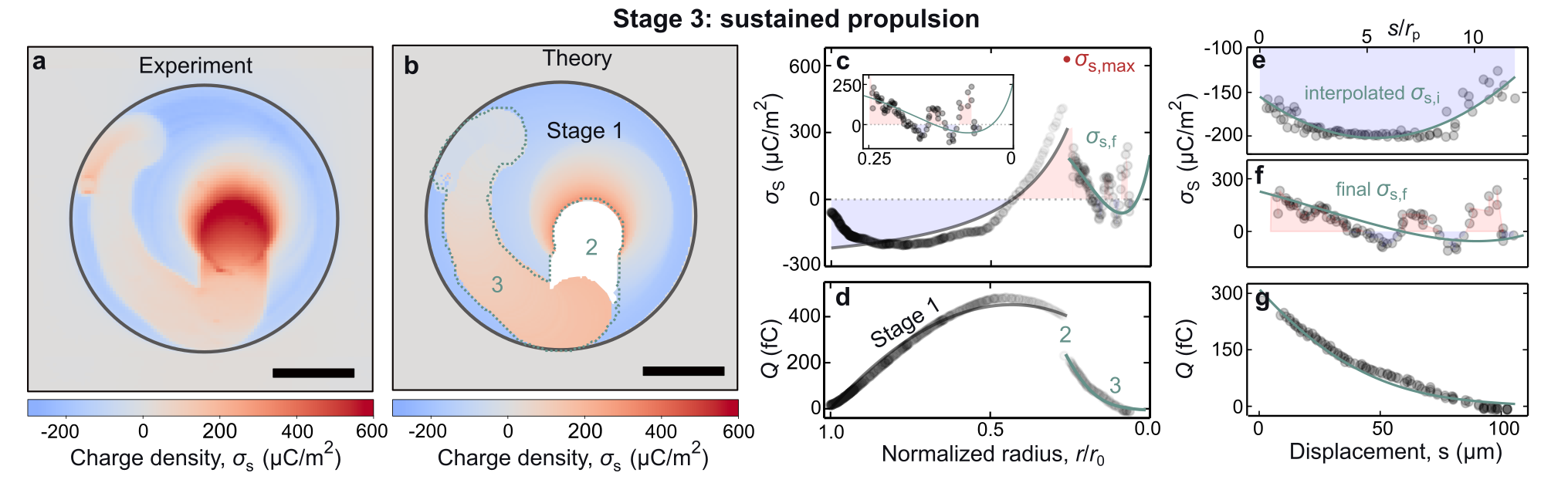}
\caption{\label{fig:propulsion} \textbf{Comparison of experimental and theoretical surface charge maps}. (a) Experimental $\sigma_{\text{s}}$ map obtained by KPFM vs. (b) theoretical $\sigma_{\text{s}}$ map. Scale bars: \SI{20}{\micro\meter}. (c) $\sigma_{\text{s}}$ and (d) $Q$ as functions of $r/r_0$ for all three stages. Circles are experimental results; the gray line is the prediction from Eqs.~\eqref{eq:sigma_anal} and \eqref{eq:Q_anal}; the blue-green line is the solution to Eq.~\eqref{eq:dQds}. (e) $\sigma_{\text{s}}$ evaluated from the interpolated map in Fig.~\ref{fig:before}(b); the blue-green line is a quadratic fit. (f) $\sigma_{\text{s}}$ and (g) $Q$ as functions of $s$ during stage 3.
  }
\end{figure*}

At the onset of propulsion, the droplet deposits a maximum $\sigma_{\text{s,max}} = +\SI{625}{\micro\coulomb\per\meter\squared}$ at the receding contact line [white semicircular line, Fig.~\ref{fig:sigma_max}(a); red point, Fig.~\ref{fig:sigma_max}(b) and Fig.~\ref{fig:propulsion}(c)], considerably larger than the \SI{330}{\micro\coulomb\per\meter\squared} value predicted by Eq.~\eqref{eq:sigma_s} using the stage-1 value $\alpha = 0.32$. Matching the measured $\sigma_{\text{s,max}}$ requires a larger $\alpha = 0.6$, suggesting that charge transfer is enhanced during the rapid contact-line motion at $U \sim \SI{10}{\centi\meter\per\second}$. This corresponds to $\mathrm{Pe} \sim 100$, and the charge-transfer process is no longer quasi-static. 

However, this burst of motion lasts only for about \SI{2.5}{\milli\second}. Over a distance of roughly one droplet diameter or \SI{16}{\micro\meter} [white to gray semicircular lines, Fig.~\ref{fig:sigma_max}(a)], $\sigma_{\text{s}}$ decreases from +625 to $+\SI{238}{\micro\coulomb\per\meter\squared}$ [Fig.~\ref{fig:sigma_max}(b)], and the droplet loses 30\% of its charge, with $Q$ decreasing from 400 to \SI{280}{\femto\coulomb} [later in Fig.~\ref{fig:propulsion}(d)]. The drop in charge $\Delta Q = \SI{-120}{\femto\coulomb}$ is calculated by numerically integrating the incremental charge loss $\mathrm{d}Q$ 
 \begin{equation}
	 \mathrm{d}Q =- 2r \left[ \sigma_\mathrm{s,f}-\sigma_\mathrm{s,i} \right] \mathrm{d}s,
	  \label{eq:dQ_2}
 \end{equation}
over $s$ = 0--16 \SI{}{\micro\meter}. Here, $\sigma_\mathrm{s,f}$, the final surface charge density, is evaluated at the receding contact line from the KPFM map in Fig.~\ref{fig:sigma_max}(a), while $\sigma_\mathrm{s,i}$, the interpolated surface charge density, is evaluated at the advancing contact line from the map in Fig.~\ref{fig:before}(b).

At this point, the droplet enters the regime of sustained self-propulsion with a much lower $U$, the subject of the next section.

\section{Stage 3: sustained propulsion}

During stage 3, the droplet self-propels at a roughly constant speed $U \approx \SI{0.18}{\milli\meter\per\second}$ [Fig.~\ref{fig:phenomenon}(c)], implying that $\tilde{F}_{e}$ and $\tilde{F}_{\text{hys}}$ are approximately matched throughout. The charge-balance formulation in Section~\ref{sec:stage1} has to be modified to account for charge leaving the droplet at the receding contact line, $\sigma_\mathrm{s,f}$, and charge entering at the advancing contact line, $\sigma_\mathrm{s,i}$ [schematic in Fig.~\ref{fig:phenomenon}(h)]:
\begin{equation}
     \frac{\mathrm{d}Q}{{\mathrm{d}s}} =- 2r \left[ \sigma_\mathrm{s,f} -\sigma_\mathrm{s,i} \right].
   \label{eq:dQds}
\end{equation}
Here, $\sigma_\mathrm{s,i}(s)$ is obtained from the same interpolated map in Fig.~\ref{fig:before}(b), but evaluated along the droplet path, while $r(s) = r_{p} - 0.064\,s$ is obtained experimentally [Fig.~\ref{fig:phenomenon}(d)]. Since $\mathrm{Pe} \sim 10^{-1}$ and the charge-transfer process is again quasi-static,
\begin{equation}
	    \sigma_\mathrm{s, f}(s) = \alpha \left( \sigma_\mathrm{EDL} + \frac{Q}{\pi r^{2}}\right).
 \label{eq:sigma_sf}
 \end{equation}
In general, Eqs.~\eqref{eq:dQds} and \eqref{eq:sigma_sf} have no analytic solutions and must be solved numerically.
 
Together, the two equations describe the coupling between surface charge deposition, droplet self-charging, and self-propulsion. We compare their predictions with experimental KPFM data [Fig.~\ref{fig:propulsion}(a)]. In Fig.~\ref{fig:propulsion}(b), the theoretical $\sigma_{\text{s}}$ map is generated using Eqs.~\eqref{eq:sigma_anal} and \eqref{eq:Q_anal} for stage 1 and Eqs.~\eqref{eq:dQds} and \eqref{eq:sigma_sf} for stage 3; stage 2 is left blank, as it cannot be predicted directly. Importantly, we use the same parameter values $\alpha = 0.32$ and $\sigma_{\mathrm{EDL}} = \SI{-665}{\micro\coulomb\per\meter\squared}$ for stages 1 and 3, and the composite theoretical map shows good quantitative agreement with experiment [Fig.~\ref{fig:propulsion}(c),(d)].

For example, Eqs.~\eqref{eq:dQds} and \eqref{eq:sigma_sf} correctly predict the emergence of a mosaic surface charge pattern [inset in Fig.~\ref{fig:propulsion}(c); Fig.~\ref{fig:propulsion}(f)]. As the droplet self-propels, it leaves behind $\sigma_{\text{s,f}}$ that alternates between polarities; this is a direct result of the non-uniform $\sigma_{\text{s,i}}$, which we fit with a quadratic function [Fig.~\ref{fig:propulsion}(e)]. A uniform $\sigma_{\text{s,i}}$ would instead result in a monotonic decrease of $\sigma_{\text{s,f}}$ [Fig.~\ref{fig:sigma_si_fit}]. Reproducing the exact oscillations measured in $\sigma_{\text{s,f}}(s)$ would likely require a different functional form for $\sigma_{\text{s,i}}(s)$ than the quadratic used here. Despite the oscillations in $\sigma_{\text{s,f}}$, the droplet charge $Q$ itself falls monotonically with $s$ [Fig.~\ref{fig:propulsion}(d),(g)].

In principle, the exact trajectory of the droplet could be predicted self-consistently by numerically integrating $\mathrm{d}\tilde{F}_{e}$ over the evolving charge landscape, analogous to how $\tilde{F}_{e}(\tilde{r})$ was obtained in Eq.~\eqref{eq:Fe} for stage 2. However, it is difficult to account for spatial variations in local pinning forces. Hence, we restrict ourselves to a qualitative discussion of this motion. 

Because the droplet carries a positive charge $Q$, it is electrostatically attracted toward regions of negative surface charge density and repelled from regions of positive surface charge density. This explains why the droplet avoids its previous path, where it just deposited positive $\sigma_{\text{s,f}}$, while seeking out regions of negative $\sigma_{\text{s,i}}$ [schematic in Fig.~\ref{fig:phenomenon}(h)]. As a result, the droplet trajectory is self-avoiding while remaining confined within its own initial footprint, where negatively charged deposits are available for recombination.

The coupling between self-charging and self-propulsion results in complex trajectories as shown in Fig.~\ref{fig:other_motion}. An even richer set of trajectories emerges when multiple droplets evaporate in close proximity. Because each droplet leaves behind its own charge landscape, a neighboring droplet can be attracted into, and propelled by, the charge deposits left by another, giving rise to not just self-avoidance but mutual avoidance of paths [Fig.~\ref{fig:multiple}].

\begin{figure*}[!htb]
\centering
\includegraphics[scale=1]{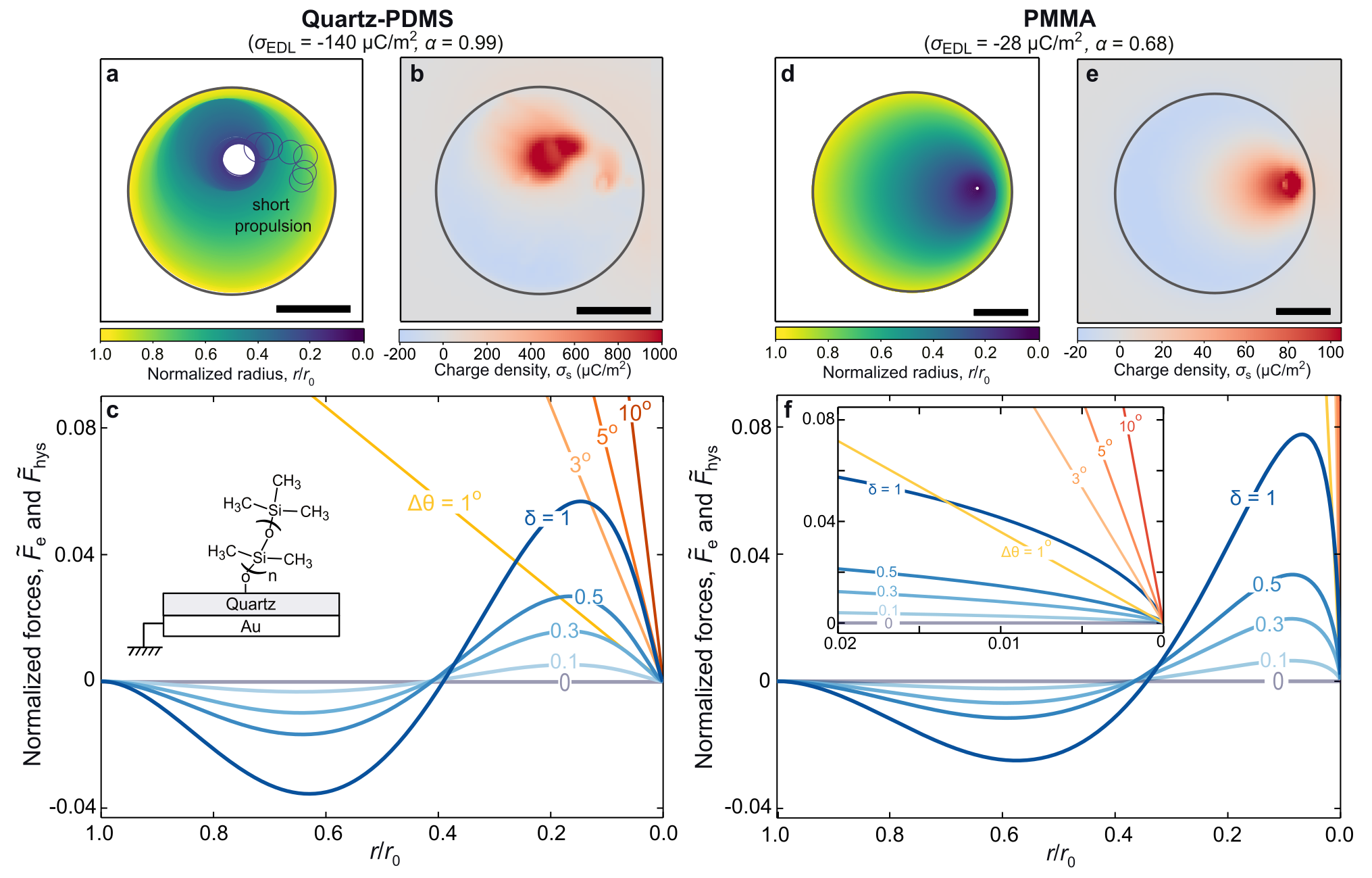}
\caption{\label{fig:other_surfaces} \textbf{Universality across surfaces.} (a) Contact-line map and (b) KPFM charge map for a droplet evaporating on quartz-PDMS ($\sigma_{\text{EDL}} = \SI{-140}{\micro\coulomb\per\meter\squared}$, $\alpha = 0.99$), showing a short self-propulsion event. (c) Corresponding $\tilde{F}_{e}$ (blue) and $\tilde{F}_{\text{hys}}$ (orange) vs.\ $\tilde{r}$; inset shows the PDMS-functionalized quartz surface. (d) Contact-line map and (e) KPFM charge map for a droplet evaporating on PMMA ($\sigma_{\text{EDL}} = \SI{-28}{\micro\coulomb\per\meter\squared}$, $\alpha = 0.68$), on which self-propulsion is never observed. (f) Corresponding $\tilde{F}_{e}$ and $\tilde{F}_{\text{hys}}$; inset shows a zoomed view near $\tilde{r} = 0$. Scale bars: \SI{20}{\micro\meter}.
  }
\end{figure*}

\section{Universality of self-propulsion}

Self-propulsion is routinely observed for droplets evaporating on fluoropolymer; the same phenomenon was reported previously by Cha \textit{et al.}~(2025), but they did not appreciate the electrostatic origin of self-propulsion \cite{cha2025pinning}. Fluoropolymers charge easily \cite{Yatsuzuka.1994}, with a large $|\sigma_{\text{EDL}}| = \SI{665}{\micro\coulomb\per\meter\squared}$.  However, this is not the full story: the charge-transfer coefficient $\alpha$ plays an equally important role in promoting self-propulsion.

To understand the subtle effects of $\alpha$, we first consider the asymptotic limits of surface charge density and droplet charge in Eqs.~\eqref{eq:sigma_anal} and \eqref{eq:Q_anal}. As $\tilde{r} \to 0$, the droplet charge vanishes as $\tilde{Q} \sim \tilde{r}^{\,2\alpha}$, but its deposited charge density $\tilde{\sigma}_{\text{s}} \sim \tilde{r}^{\,2\alpha-2}$ diverges. The force acting on the droplet is dominated by charge deposited at radii comparable to $\tilde{r}$, which carries a charge $\sim \tilde{\sigma}_{\text{s}}\tilde{r}^{2}$ at a separation $\sim\tilde{r}$, so applying Coulomb's law we expect
\begin{equation}
\begin{split}
	\tilde{F}_{e} \sim \frac{\tilde{Q}\left(\tilde{\sigma}_{\text{s}}\tilde{r}^{2}\right)}{\tilde{r}^{2}} = \tilde{Q}\,\tilde{\sigma}_{\text{s}} \sim \tilde{r}^{\,4\alpha-2},
\end{split}
\label{eq:Fe_scaling}
\end{equation}
with a threshold at $\alpha = 1/2$ where the exponent changes sign [Appendix~\ref{app:asym}]. Eq.~\eqref{eq:Fe_scaling} explains why $\tilde{F}_{e}(\tilde{r})$ diverges as $\tilde{r} \to 0$ for the fluoropolymer with $\alpha = 0.32$ [Fig.~\ref{fig:onset}(b)]. This divergence is crucial because it means that $\tilde{F}_{e}$ eventually exceeds the pinning force $\tilde{F}_{\text{hys}}$, irrespective of contact angle hysteresis $\Delta\theta$.

By contrast, for $\alpha > 1/2$, $\tilde{F}_{e} \to 0$ as $\tilde{r} \to 0$ [Eq.~\eqref{eq:Fe_scaling}]; whether $\tilde{F}_{e}$ intersects $\tilde{F}_{\text{hys}}$ depends sensitively on $\Delta\theta$ and $\sigma_{\text{EDL}}$. We illustrate this point with two surfaces both having $\alpha > 1/2$: a \SI{0.3}{\micro\meter}-thick amorphous quartz film grafted with polydimethylsiloxane brushes (Quartz-PDMS) and a \SI{4}{\micro\meter}-thick film of polymethylmethacrylate (PMMA). Quartz-PDMS has $\alpha = 0.99$ and $\sigma_{\text{EDL}} = \SI{-140}{\micro\coulomb\per\meter\squared}$, while PMMA has $\alpha = 0.68$ and $\sigma_{\text{EDL}} = \SI{-28}{\micro\coulomb\per\meter\squared}$ [Fig.~\ref{fig:other_surfaces}]. The values used here were determined using the method described in Section \ref{sec:stage1}.

On quartz-PDMS, self-propulsion is observed only occasionally, and when it does occur, the resulting trajectory is short [Fig.~\ref{fig:other_surfaces}(a),(b)]. On PMMA, by contrast, self-propulsion is never observed even when the retraction dynamics are asymmetrical [Fig.~\ref{fig:other_surfaces}(d),(e)]. This difference is readily explained by the corresponding $\tilde{F}_{e}$ and $\tilde{F}_{\text{hys}}$ curves, calculated using Eqs.~(\ref{eq:Fe}) and (\ref{eq:Fhys}) [Fig.~\ref{fig:other_surfaces}(c),(f)]. Here, we use $\varepsilon_{s}$ = 3.7 for both quartz and PMMA.

On quartz-PDMS, because of the relatively large $|\sigma_{\text{EDL}}|$, the $\tilde{F}_{e}$ and $\tilde{F}_{\text{hys}}$ curves can intersect as long as $\Delta\theta < 5^\circ$ [Fig.~\ref{fig:other_surfaces}(c)]. For example, when $\Delta\theta = 3^\circ$ and $\delta = 1$, $\tilde{F}_{e}$ intersects $\tilde{F}_{\text{hys}}$ at $\tilde{r}_{\text{p}} = 0.13$, a physically reasonable number. By contrast, for PMMA, with its much smaller $|\sigma_{\text{EDL}}|$, $\tilde{F}_{e}$ and $\tilde{F}_{\text{hys}}$ intersect only at $\tilde{r}_{\text{p}} = 0.001$ for the same $\Delta\theta = 3^\circ$ and $\delta = 1$ [inset in Fig.~\ref{fig:other_surfaces}(f)], i.e., only after the droplet has shrunk to less than 0.1\% of its initial footprint, a requirement so demanding that self-propulsion is practically never observed.

\section{Conclusion}

In summary, we show that droplet self-propulsion is electrostatic in origin. The coupling between droplet self-charging and self-propulsion produces mosaic charge patterns and self-avoiding trajectories that displace droplets by many times their own diameter. We further identify the crucial role played by the electric double layer, specifically the charge-transfer coefficient $\alpha$, in setting the onset of propulsion: for $\alpha < 1/2$, self-propulsion is strongly favored, whereas for $\alpha > 1/2$ whether a droplet propels depends on a delicate interplay between $\alpha$, $\sigma_{\text{EDL}}$, and $\Delta \theta$. The final fate of an evaporating sessile droplet is therefore shaped by the electric double layer, an interfacial region orders of magnitude thinner than the droplet it sets in motion.

\section{Materials and Methods \label{sec:methods}}

Most of the materials and methods have been described in a previous manuscript \cite{Bipolar_singh_2025}, and are reproduced here for completeness. \bigskip

\noindent \textbf{Materials.}
Polydimethylsiloxane (PDMS, 1000\,cP) and deionized water (HPLC grade) were purchased from Sigma-Aldrich. Polymethylmethacrylate (PMMA) sheet was purchased from Rowad Plastic. 
\bigskip

\begin{figure}[!htb]
 \centering
 \includegraphics[scale=1]{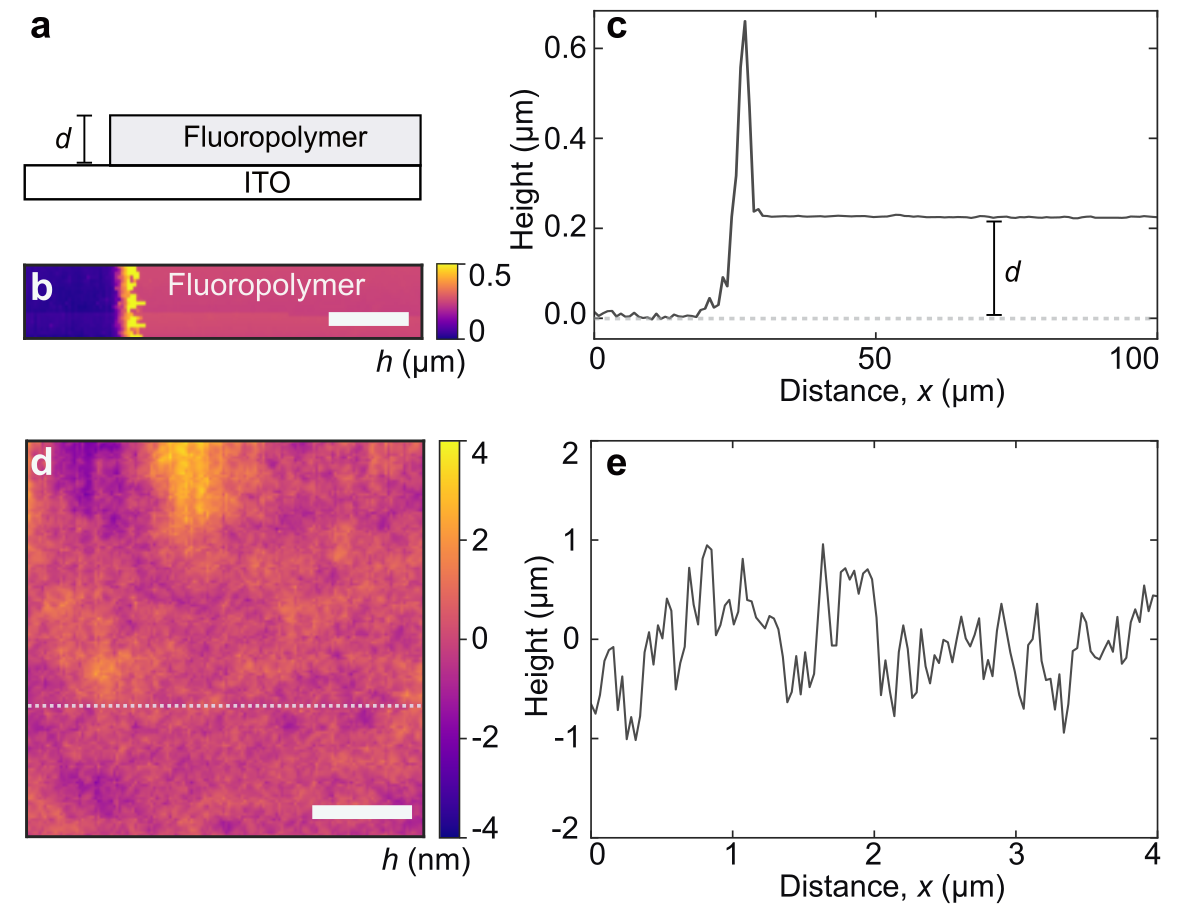}
 \caption{\label{fig:surface}\textbf{Surface characterization.} (a) Fluoropolymer film of thickness $d$ was deposited on an ITO substrate. (b) AFM micrograph (scale bar: \SI{20}{\micro\meter}) and (c) height profile across a scratched region, used to determine $d$. (d) AFM micrograph (scale bar: \SI{1}{\micro\meter}) and (e) corresponding height profile reveals a surface roughness of about \SI{1}{\nano\meter}.}
 \end{figure}
 
\noindent \textbf{Fluoropolymer film.} A thin fluoropolymer film of thickness $d$ was deposited onto an ITO-coated glass substrate by C$_4$F$_8$ plasma-enhanced chemical vapour deposition~\cite{terriza2014ppapC4F8} [Fig.~\ref{fig:surface}(a)].

A small region of the film was mechanically scratched down to the underlying ITO layer and imaged by atomic force microscopy (AFM) to reveal a thickness of $d = \SI{220}{\nano\meter}$ [Fig.~\ref{fig:surface}(b)]. AFM imaging of an unscratched region [Fig.~\ref{fig:surface}(d)] confirms that the film is smooth and homogeneous (with roughness of $\sim\SI{1}{\nano\meter}$) over the micrometer length scales relevant to contact-line pinning [Fig.~\ref{fig:surface}(e)].

\bigskip 
\noindent \textbf{PMMA film.} The PMMA sheet was cut into small pieces, mixed with toluene, and stirred for 4\,h at room temperature to obtain a uniform solution. The solution (5\,wt\% PMMA) was filtered through a hydrophobic PTFE syringe filter (\SI{0.22}{\micro\metre} pore size) before being spin-coated onto an ITO-coated glass slide at 500\,rpm for a few seconds to yield a \SI{4}{\micro\metre}-thick film \cite{hall1998spin}.
 
\bigskip
\noindent \textbf{Quartz-PDMS film.} A 300 nm-thick \ce{SiO2} (amorphous quartz) layer was deposited on an Au/Ti-coated glass slide using Plasma Enhanced Chemical Vapor Deposition (PECVD, Oxford Plasma Lab System 100) at \SI{300}{\degreeCelsius} and a deposition rate of \SI{60}{nm/min}. The quartz surface was then thoroughly cleaned with IPA and plasma oxidized for \SI{90}{s}. We follow the recipe by Wang, McCarthy, and co-workers to functionalize the surface \cite{wang2016covalently}. Immediately after plasma cleaning, the quartz surface was dipped into a solution containing 10\,wt\% of PDMS (1000\,cP) and 1\,wt\% H$_2$SO$_4$ in IPA. The surface was promptly withdrawn, placed on a tilted holder, and allowed to react for 12\,h. Following the reaction, the surface was thoroughly rinsed sequentially with toluene, IPA, and water.
 
\bigskip
\noindent \textbf{Generating droplets.}
A commercially available \SI{100}{mL} PET spray bottle was first rinsed with Millipore water and ethanol. The entire bottle, including the nozzle and the spray pipe, was submerged in 70\% (v/v) ethanol in water and sonicated for \SI{30}{min} to remove residual chemicals. The bottle was then rinsed again with Millipore water, wrapped in aluminium foil, and heated in a drying oven at \SI{55}{\degreeCelsius} overnight. Finally, the bottle was filled with deionized, HPLC-grade water for spraying.

\bigskip
\begin{figure}[!htb]
    \centering
    \includegraphics[scale=0.95]{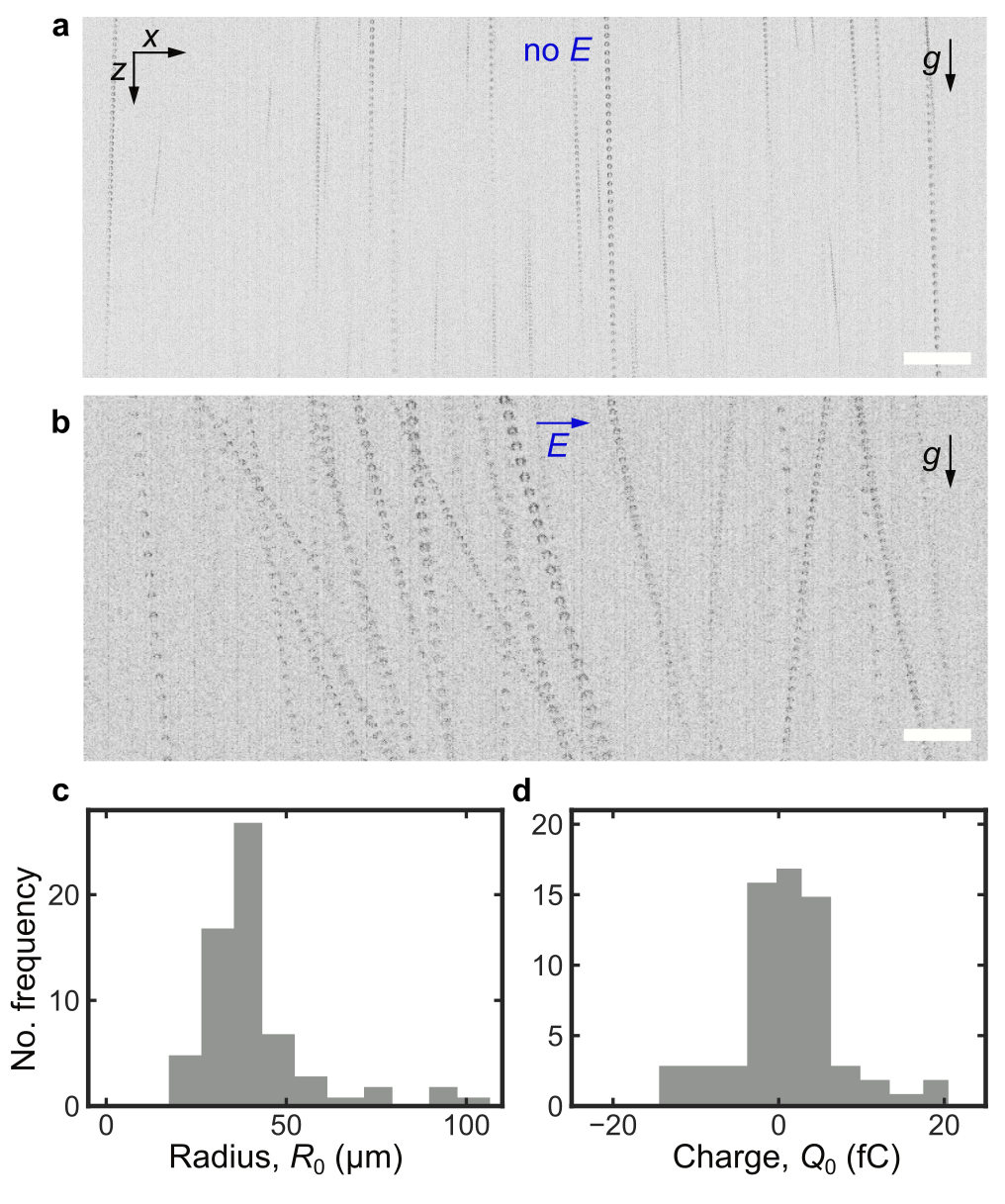}
    \caption{\textbf{Measuring droplet size and charge.} Trajectory of sprayed droplets (a) under gravity alone and (b) in the presence of electric field $E = \SI{8e4}{\volt\per\meter}$. Scale bars: 0.5 mm. Time interval between snapshots is 0.7 ms. Histograms of droplet (c) radii and (d) electrical charges obtained by applying Eqs.~\eqref{eq:r} and \eqref{eq:q}, respectively (n = 65 droplets).
    }
    \label{fig:distribution}
\end{figure}

\noindent \textbf{Measuring droplet size and charge.} Droplets were sprayed from a height $\sim$ \SI{20}{\centi\meter} above the substrate.  Before landing on the substrate, the droplets passed through two parallel aluminium plates (\SI{8}{\centi\meter} $\times$ \SI{8}{\centi\meter}, spaced \SI{2.5}{\centi\meter} apart) with a \SI{2}{\kilo\volt} potential applied across, corresponding to an electric field $E = \SI{8e4}{\volt\per\meter}$.

In the absence of an applied field, droplets followed straight vertical trajectories with terminal velocity $U_{z}$ [Fig.~\ref{fig:distribution}(a)]. Balancing the droplet's weight $\tfrac{4}{3}\pi R_{0}^{3}\rho g$ against the Stokes drag $6\pi \eta R_{0} U_{z}$ yields the droplet radius
\begin{equation} \label{eq:r}
    R_{0} = \left( \frac{9 \eta U_{z}}{2 \rho g} \right)^{1/2},
\end{equation}
where $\rho$ is the liquid density and $\eta$ the air viscosity.

In the presence of an applied field, droplets were deflected laterally and moved with a horizontal velocity $U_{x}$: positively charged droplets in the direction of the field [rightward in Fig.~\ref{fig:distribution}(b)], and negatively charged droplets in the opposite direction. Balancing the electrical force $Q_{0}E$ against the Stokes drag $6\pi \eta R_{0} U_{x}$ yields the droplet charge
\begin{equation} \label{eq:q}
    Q_{0} = \frac{6\pi \eta\, U_{x}}{E} \left( \frac{9\,\eta\,U_{z}}{2\rho g} \right)^{1/2}.
\end{equation}

Across a total of 65 sprayed droplets analyzed, the average radius calculated using Eq.~\eqref{eq:r} was $R_{0}$ = 42 $\pm$ 17\,\SI{}{\micro\meter} [Fig.~\ref{fig:distribution}(c)]. Similarly, using Eq.~\eqref{eq:q}, the calculated droplet charge $Q_{0}$ ranged from $-15$ to $+21\,$\SI{}{\femto\coulomb} and was roughly centered on zero with $\langle Q_{0} \rangle$ = 1 $\pm$ 7\,\SI{}{\femto\coulomb} [Fig.~\ref{fig:distribution}(d)].

By the time the droplets landed on the substrate, they had further evaporated during their descent, reducing their size below the radius $R_0$ measured in flight. The initial footprint radius upon contact $r_0 = 29\pm7\,$\SI{}{\micro\meter} was determined using an optical microscope.

\bigskip
\noindent \textbf{Kelvin Probe Force Microscopy.} KPFM measurements were performed in dual-pass mode 
on a Bruker JPK NanoWizard 5 using a Pt-Ir-coated conductive tip (Bruker SCM-PIT-V2; spring constant $k \sim 5~\text{N\,m}^{-1}$, tip radius $\sim 5~\text{nm}$, resonant frequency $\sim 75~\text{kHz}$). In the first pass, the tip scans the surface in tapping mode to record the topography. In the second pass, the tip retraces the same line at a constant lift height of tens of nanometers above the recorded profile, with the mechanical excitation switched off and an AC voltage bias $V_{\text{ac}}$ applied instead. Operating in lift mode decouples the long-range electrostatic interaction from short-range interactions such as van der Waals and capillary forces. The local contact potential difference is given by the DC offset voltage $V_{\text{dc}}$ that nullifies the electrically driven cantilever oscillation, as determined by a feedback loop [see schematic in Fig.~\ref{fig:phenomenon}(g)].

The surface charge density $\sigma_{\text{s}}$ follows from a parallel-plate capacitor model,
\begin{equation}
    \sigma_{\text{s}} = \frac{\varepsilon_{0}\varepsilon_{\text{s}} \Delta V_{\text{dc}}}{d},
\label{eq:sigma_abs}
\end{equation}
where $\Delta V_{\text{dc}}$ is the contact potential difference relative to the uncharged surface outside the droplet footprint; $d$ is the substrate thickness; and $\varepsilon_{\text{s}}$ its relative dielectric permittivity. This relation holds when the surface charge varies slowly over the lateral scale of the substrate thickness, i.e., $|\sigma_{\text{s}}/\nabla\sigma_{\text{s}}| \gg d$, which is satisfied everywhere in our measurements except at the footprint edge.

\bigskip

\section*{Acknowledgments}
A.D.R.\ is supported by the European Union's Horizon 2020 research and innovation program (grant agreement no.\ 883631). D.D.\ acknowledges support from JSPS KAKENHI (Grant number JP26K24707) and from the Okinawa Institute of Science and Technology Graduate University (OIST).
  
\section*{Author contributions}
D.D.\ proposed the work and supervised the research; N.S.\ designed, conducted, and evaluated the experiments; D.D., A.D.R., and N.S. jointly developed the analytical model; S.N.R.\ Kazmi prepared the fluoropolymer surface;  D.D., N.S., and A.D.R.\ jointly interpreted the results and prepared the manuscript.

\section*{Data availability} Data and computer codes supporting the findings of this article are openly available \cite{dataset}.

\section*{Appendices}
\appendix

\section{Fast motion at onset of propulsion}
\label{app:fast_motion}
\begin{figure}[!htb]
    \centering
    \includegraphics[scale=1]{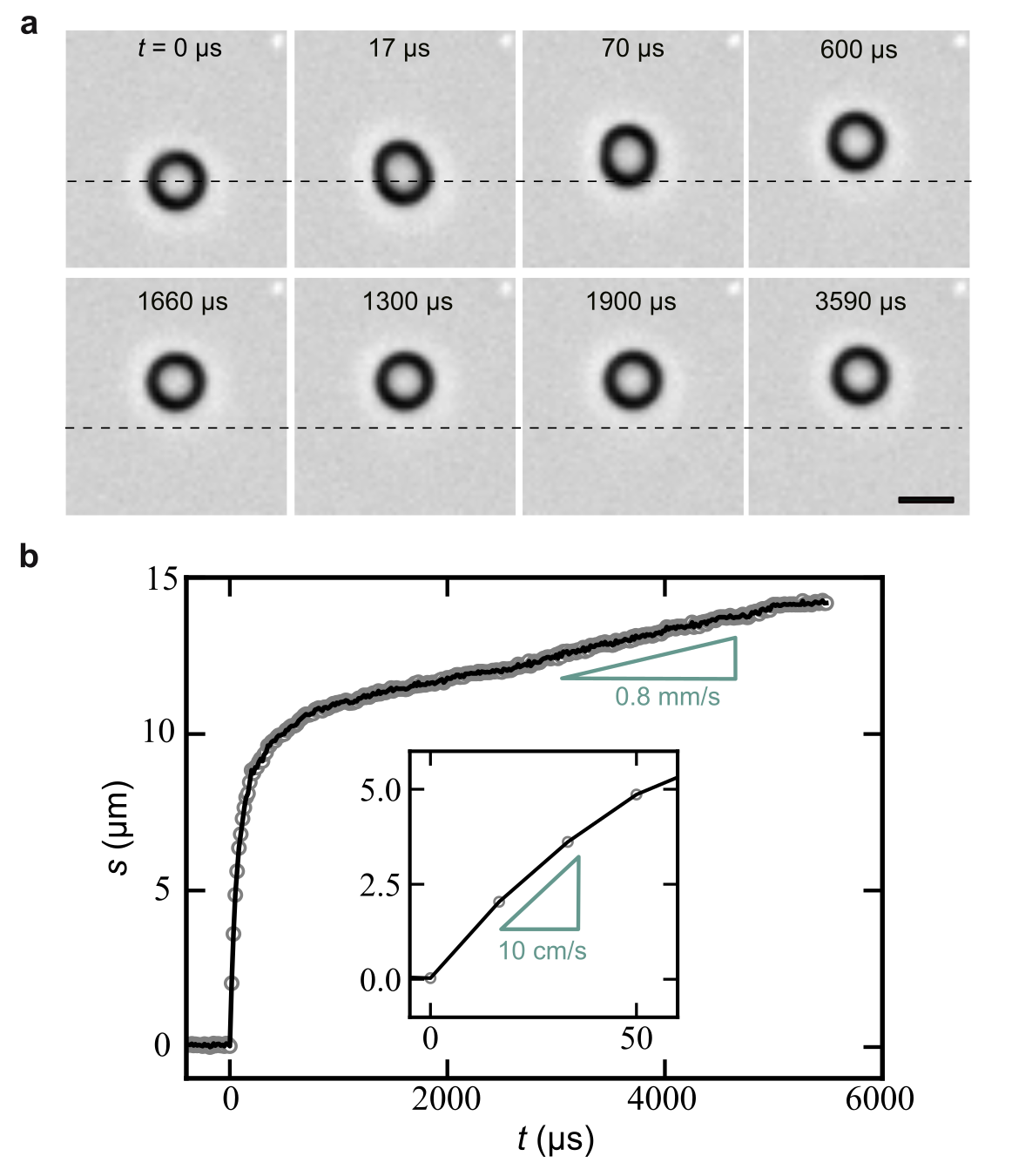}
    \caption{\textbf{Fast propulsion during stage 2.} (a) Timelapse images at the onset of propulsion. Scale bar: \SI{15}{\micro\meter}. (b) Corresponding displacement $s$ vs.\ time $t$. Inset shows that the droplet reaches peak velocity of $\sim \SI{10}{\centi\meter\per\second}$.
    }
    \label{fig:fast}
\end{figure}

\section{Closed form of $\mathcal{G}$}
\label{app:G}
The geometric kernel $\mathcal{G}$ in Eq.~\eqref{eq:G} can be evaluated in closed form. Substituting $\varphi = 2\psi$ in Eq.~\eqref{eq:L} brings $\tilde{L}$ to Legendre normal form,
\begin{equation}
	\tilde{L}^{2} = (\Delta\tilde{x}+\tilde{r}')^{2}\left(1-m\sin^{2}\psi\right),
\label{eq:legendre}
\end{equation}
where
\begin{equation}
	m = \frac{4\Delta\tilde{x}\,\tilde{r}'}{(\Delta\tilde{x}+\tilde{r}')^{2}}
\label{eq:modulus}
\end{equation}
is the elliptic parameter, which measures how off-centre the droplet sits relative to the ring.

The numerator $(1-\delta\cos\varphi) (\Delta\tilde{x}-\tilde{r}'\cos\varphi)$ in $\mathcal{G}$ can be expanded as  $ \Delta\tilde{x} - (\tilde{r}'+\delta\Delta\tilde{x})\cos\varphi + \delta\tilde{r}'\cos^{2}\varphi$, i.e., polynomials in $\cos\varphi$. Therefore, $\mathcal{G}$ is a weighted sum of three angular moments of $\tilde{L}^{-3}$. Each moment can in turn be written in terms of the complete elliptic integrals $K(m)$ and $E(m)$ of the first and second kind \cite{DLMF, GR2007}. For $\mathcal{I}_{n} = \int_{0}^{\pi/2}\sin^{n}\!\psi\,(1-m\sin^{2}\psi)^{-3/2}\,\mathrm{d}\psi$ and $\mathcal{P} = 4/(\Delta\tilde{x}+\tilde{r}')^{3}$, the three moments are then
\begin{equation}
\begin{split}
	\int_{0}^{2\pi}\!\frac{\mathrm{d}\varphi}{\tilde{L}^{3}} &= \mathcal{P}\,\mathcal{I}_{0}, \\
	\int_{0}^{2\pi}\!\frac{\cos\varphi\,\mathrm{d}\varphi}{\tilde{L}^{3}} &= \mathcal{P}\left(2\mathcal{I}_{2}-\mathcal{I}_{0}\right), \\
	\int_{0}^{2\pi}\!\frac{\cos^{2}\varphi\,\mathrm{d}\varphi}{\tilde{L}^{3}} &= \mathcal{P}\left(4\mathcal{I}_{4}-4\mathcal{I}_{2}+\mathcal{I}_{0}\right),
\end{split}
\label{eq:moments}
\end{equation}
with
\begin{equation}
\begin{split}
	\mathcal{I}_{0} &= \frac{E(m)}{1-m}, \qquad \mathcal{I}_{2} = \frac{\mathcal{I}_{0}-K(m)}{m}, \\
	\mathcal{I}_{4} &= \frac{\mathcal{I}_{0}-2K(m)+E(m)}{m^{2}}.
\end{split}
\label{eq:In}
\end{equation}
Collecting terms,
\begin{equation}
\begin{split}
	\mathcal{G} = \mathcal{P}\Big[ &\Delta\tilde{x}\,\mathcal{I}_{0} - (\tilde{r}'+\delta\Delta\tilde{x})(2\mathcal{I}_{2}-\mathcal{I}_{0}) \\
	&+ \delta\,\tilde{r}'(4\mathcal{I}_{4}-4\mathcal{I}_{2}+\mathcal{I}_{0}) \Big].
\end{split}
\label{eq:G_closed}
\end{equation}
Since $\mathcal{G}$ can be evaluated analytically, Eq.~\eqref{eq:Fe} requires only a single numerical integration over $\tilde{r}'$.

\section{Characteristics of $\tilde{F}_{e}(\tilde{r})$ for $\alpha = 0.32$}

The critical radius $\tilde{r}_{\text{crit}}$, at which $\tilde{F}_{e}$ changes sign, depends only weakly on the droplet shift $\delta$ [Table~\ref{tab:rcrit}]. It varies by about 10\% over the entire range $0 < \delta \leq 1$. The radius $\tilde{r}_{\text{min}}$ at which the attractive force is strongest is similarly insensitive to $\delta$, and always exceeds $\tilde{r}_{\text{rev}} = 0.43$. By contrast, the depth of the attractive well $\tilde{F}_{e}(\tilde{r}_{\text{min}})$ scales nearly linearly with $\delta$, vanishing as $\delta \to 0$ as required by symmetry.

\begin{table}[!htb]
\centering
\caption{\textbf{Characteristic radii of $\tilde{F}_{e}(\tilde{r})$ for $\alpha = 0.32$.} $\tilde{r}_{\text{crit}}$ is the radius at which $\tilde{F}_{e}$ changes sign, and $\tilde{r}_{\text{min}}$ the radius at which it is most attractive.}
\label{tab:rcrit}
\begin{ruledtabular}
\begin{tabular}{cccc}
$\delta$ & $\tilde{r}_{\text{crit}}$ & $\tilde{r}_{\text{min}}$ & $\tilde{F}_{e}(\tilde{r}_{\text{min}})$ \\
\hline
0.05 & 0.2827 & 0.4973 & $-0.0005$ \\
0.10 & 0.2826 & 0.4973 & $-0.0010$ \\
0.20 & 0.2820 & 0.4955 & $-0.0020$ \\
0.30 & 0.2811 & 0.4955 & $-0.0030$ \\
0.50 & 0.2779 & 0.4901 & $-0.0052$ \\
0.70 & 0.2722 & 0.4829 & $-0.0075$ \\
0.80 & 0.2680 & 0.4775 & $-0.0087$ \\
0.90 & 0.2623 & 0.4720 & $-0.0101$ \\
1.00 & 0.2543 & 0.4630 & $-0.0116$ \\
\end{tabular}
\end{ruledtabular}
\end{table}

\section{Asymptotics for $\tilde{F}_{e}(\tilde{r})$}
\label{app:asym}

As $\tilde{r} \to 0$ (for $0 < \alpha < 1$), Eqs.~\eqref{eq:sigma_anal} and \eqref{eq:Q_anal} simplify to 
\begin{equation}
\begin{split}
	\tilde{\sigma}_{\text{s}}(\tilde{r}) &\sim -\frac{\alpha^{2}}{1-\alpha}\,\tilde{r}^{\,2\alpha-2}, \\
	\tilde{Q}(\tilde{r}) &\sim -\frac{\alpha}{1-\alpha}\,\tilde{r}^{\,2\alpha}.
\end{split}
\label{eq:sigma_Q_asym}
\end{equation}

In this limit, $\tilde{F}_{e}$ is dominated by rings deposited at radii $\tilde{r}'$ comparable to the droplet radius $\tilde{r}$. It is therefore natural to measure the deposition radius in units of the current one, $\tilde{r}' = u\tilde{r}$. Both lengths in the ring geometry then scale with $\tilde{r}$,
\begin{equation}
\begin{split}
	\Delta\tilde{x} &= \tilde{r}\,g, \qquad g = \delta(u-1)-u\cos\varphi, \\
	\tilde{L} &= \tilde{r}\,\ell, \qquad \ell = \left(g^{2}+u^{2}\sin^{2}\varphi\right)^{1/2},
\end{split}
\label{eq:rescaled}
\end{equation}
so that $g$ and $\ell$ are simply $\Delta\tilde{x}$ and $\tilde{L}$ expressed in units of $\tilde{r}$. Consequently the elliptic parameter of Eq.~\eqref{eq:modulus} becomes
\begin{equation}
	m = \frac{4\delta u(u-1)}{\left[\delta(u-1)+u\right]^{2}},
\label{eq:m_selfsim}
\end{equation}
which depends only on $u$ and $\delta$: shrinking the droplet rescales the entire ring configuration without deforming it. Since $\mathcal{G}$ carries dimensions of inverse length squared [Eq.~\eqref{eq:G}], it must then take the form
\begin{equation}
\begin{split}
	\mathcal{G} &= \frac{1}{\tilde{r}^{2}}\,J(u,\delta), \\
	J(u,\delta) &= \int_{0}^{2\pi}\frac{(1-\delta\cos\varphi)\,g}{\ell^{3}}\,\mathrm{d}\varphi,
\end{split}
\label{eq:J}
\end{equation}
with the whole $\tilde{r}$ dependence carried by the prefactor and none by $J$.

Inserting Eqs.~\eqref{eq:sigma_Q_asym} and \eqref{eq:J} into Eq.~\eqref{eq:Fe} with $\mathrm{d}\tilde{r}' = \tilde{r}\,\mathrm{d}u$, all powers of $\tilde{r}$ collect into a single factor and the $u$-integral reduces to a pure number,
\begin{equation}
\begin{split}
	\tilde{F}_{e}(\tilde{r}) &\sim \left[ \frac{2}{1 + \varepsilon_{s}} - \frac{1}{\varepsilon_{s}} \right] \frac{\alpha^{3}}{(1-\alpha)^{2}}\,C(\delta,\alpha)\;\tilde{r}^{\,4\alpha-2}, \\
	C(\delta,\alpha) &= \int_{1}^{\infty} u^{2\alpha-1} J(u,\delta)\,\mathrm{d}u.
\end{split}
\label{eq:Fe_asym}
\end{equation}
The upper limit of $C(\delta,\alpha)$ deserves a comment. Two well-separated rings behave as $J \to 2/u^{2}$, so the integrand $u^{2\alpha-1}J$ decays as $u^{2\alpha-3}$, which is integrable at infinity for every $\alpha < 1$. The integral therefore converges: charge deposited early, when the droplet was much larger, contributes to the amplitude $C$ but never to the exponent.

This scaling for $\tilde{F}_{e}$ has a threshold at $\alpha = 1/2$, where the exponent changes sign. For $\alpha < 1/2$, $\tilde{F}_{e}$ diverges as $\tilde{r} \to 0$, so it always eventually exceeds the pinning force $\tilde{F}_{\text{hys}}$, which vanishes linearly with $\tilde{r}$ [Eq.~\eqref{eq:Fhys}], regardless of $\Delta\theta$. By contrast, for $\alpha > 1/2$, $\tilde{F}_{e} \to 0$ as $\tilde{r} \to 0$, so whether $\tilde{F}_{e}$ intersects $\tilde{F}_{\text{hys}}$ before both vanish depends sensitively on $\Delta\theta$ and $\sigma_{\text{EDL}}$.

The results also follow from a simple physical argument: the droplet carries $\tilde{Q} \sim \tilde{r}^{\,2\alpha}$, the ring it has just deposited carries a comparable charge $\tilde{\sigma}_{\text{s}}\tilde{r}^{2} \sim \tilde{r}^{\,2\alpha}$, and the two are separated by $\sim\tilde{r}$, so that $\tilde{F}_{e} \sim \tilde{r}^{\,2\alpha}\tilde{r}^{\,2\alpha}/\tilde{r}^{2} = \tilde{r}^{\,4\alpha-2}$. What Eq.~\eqref{eq:Fe_asym} adds is the geometric coefficient $C(\delta,\alpha)$, which the physical argument alone leaves undetermined.

\section{Measured onset radii}
\label{app:rp_exp}


\begin{table}[!htb]
\centering
\caption{\textbf{Measured onset radii for 19 self-propelling droplets on fluoropolymer.} $r_{0}$ is the initial footprint radius and $r_{p}$ the radius at the onset of propulsion. The last row gives the mean $\pm$ one standard deviation.}
\label{tab:rp_exp}
\begin{ruledtabular}
\begin{tabular}{cccc}
Droplet & $r_{0}$ (\si{\micro\meter}) & $r_{p}$ (\si{\micro\meter}) & $\tilde{r}_{p}=r_{p}/r_{0}$ \\
\hline
d1 & 25.5 & 10.20 & 0.400 \\
d2 & 32.9 & 9.92 & 0.302 \\
d3 & 18.4 & 5.54 & 0.301 \\
d4 & 40.2 & 17.02 & 0.423 \\
d5 & 22.9 & 5.57 & 0.244 \\
d6 & 33.8 & 9.04 & 0.267 \\
d7 & 22.2 & 5.83 & 0.262 \\
d8 & 43.9 & 10.36 & 0.236 \\
d9 & 19.5 & 3.88 & 0.199 \\
d10 & 24.0 & 3.46 & 0.144 \\
d11 & 37.0 & 8.93 & 0.242 \\
d12 & 24.2 & 5.92 & 0.245 \\
d13 & 22.5 & 5.09 & 0.226 \\
d14 & 24.8 & 6.78 & 0.273 \\
d15 & 32.6 & 10.72 & 0.329 \\
d16 & 32.9 & 8.61 & 0.261 \\
d17 & 18.7 & 5.28 & 0.283 \\
d18 & 24.3 & 6.09 & 0.250 \\
d19 & 37.8 & 10.90 & 0.289 \\
\hline
Mean & $28 \pm 8$& $8 \pm 3$& $0.27 \pm 0.06$\\
\end{tabular}
\end{ruledtabular}
\end{table}

\newpage
\section{Effect of the $\sigma_{\text{s,i}}$ fitting function}

\begin{figure}[!htb]
    \centering
    \includegraphics[scale=1]{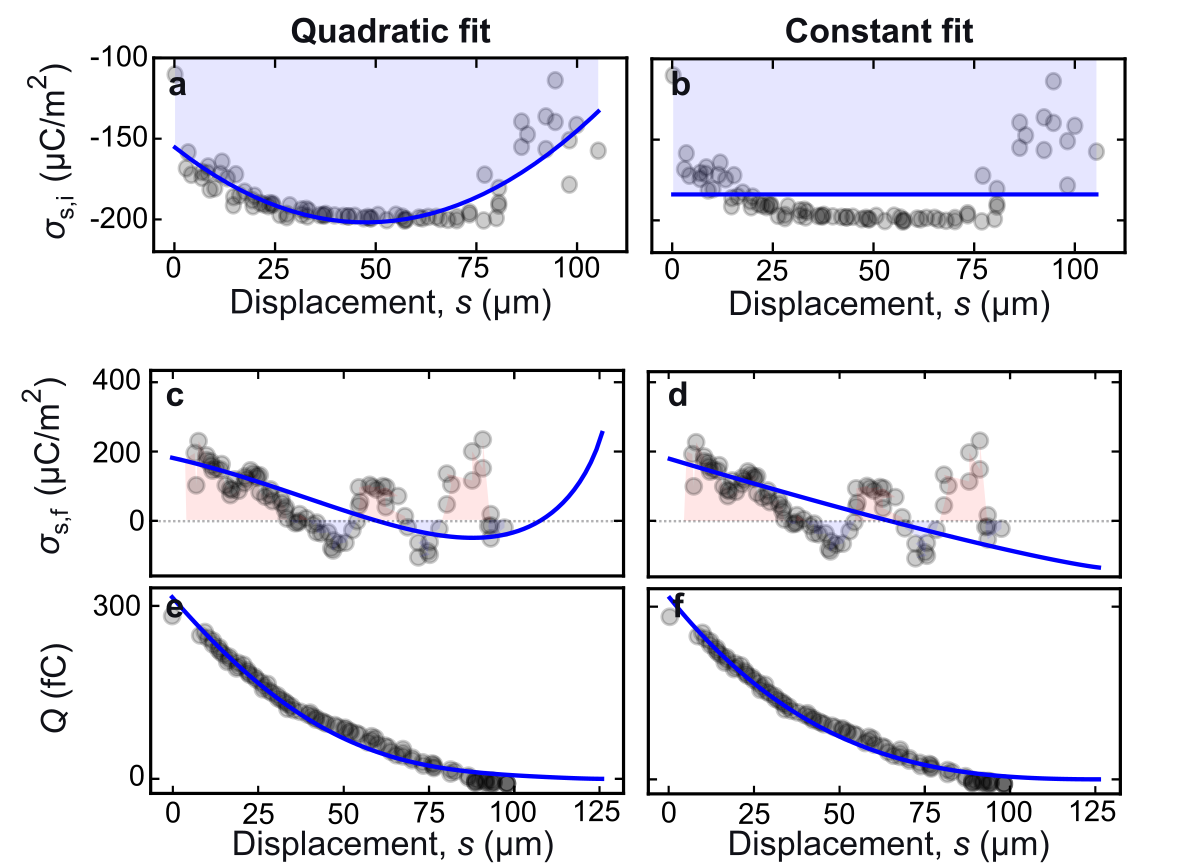}
    \caption{\textbf{Choice of fitting function for the pre-charge density.} $\sigma_{\text{s,i}}(s)$ can be fitted with either (a) a quadratic function or (b) a constant, resulting in (c) a non-monotonic and (d) a monotonically decreasing $\sigma_{\text{s,f}}(s)$, respectively. (e, f) In both cases, the predicted $Q(s)$ is monotonically decreasing. Circles: experimental data; lines: model prediction.
    }
    \label{fig:sigma_si_fit}
\end{figure}

The choice of fitting function for $\sigma_{\text{s,i}}(s)$ has an important effect on the resulting $\sigma_{\text{s,f}}(s)$. Notably, a constant fit [Fig.~\ref{fig:sigma_si_fit}(b)] results in a smooth, monotonic decay [Fig.~\ref{fig:sigma_si_fit}(d)], whereas a quadratic fit [Fig.~\ref{fig:sigma_si_fit}(a)] results in an oscillating $\sigma_{\text{s,f}}(s)$ [Fig.~\ref{fig:sigma_si_fit}(c)]. Reproducing the exact oscillations measured in $\sigma_{\text{s,f}}(s)$ would likely require a different functional form for $\sigma_{\text{s,i}}(s)$ than the quadratic used here, for instance one containing sinusoidal terms. We do not pursue this further, since $\sigma_{\text{s,i}}(s)$ is itself an interpolated quantity rather than a direct KPFM measurement. Fitting additional structure to match every oscillation risks over-interpreting features of the interpolation rather than genuine physics.

\section{Other examples}

\begin{figure}[!htb]
    \centering
    \includegraphics[scale=1]{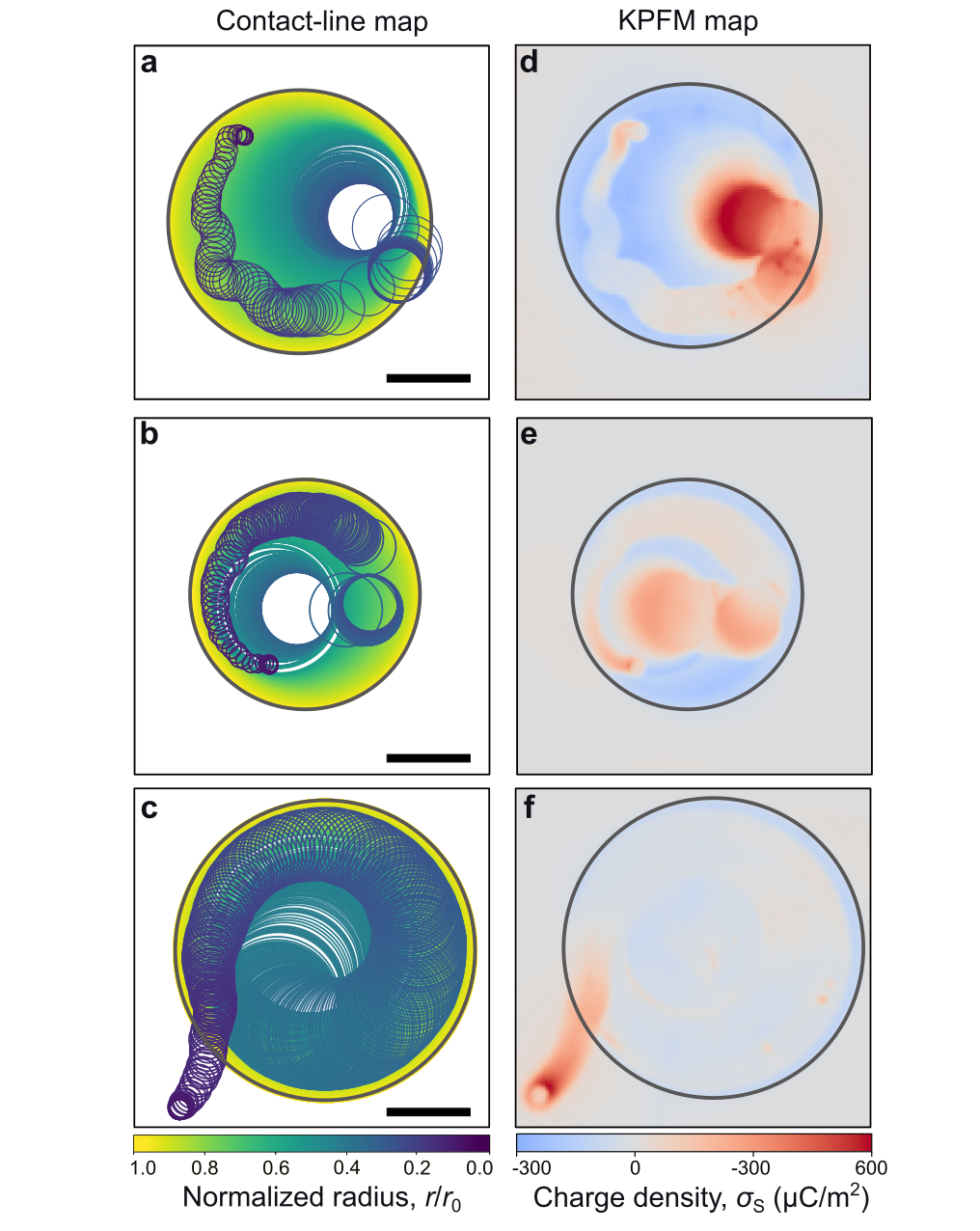}
    \caption{\textbf{Self-propulsion of single droplets.} (a--c) Contact-line maps and (d--f) corresponding KPFM maps of self-propelling droplets. Scale bars: \SI{20}{\micro\meter}.
    }
    \label{fig:other_motion}
\end{figure}

\begin{figure}[!htb]
    \centering
    \includegraphics[scale=1]{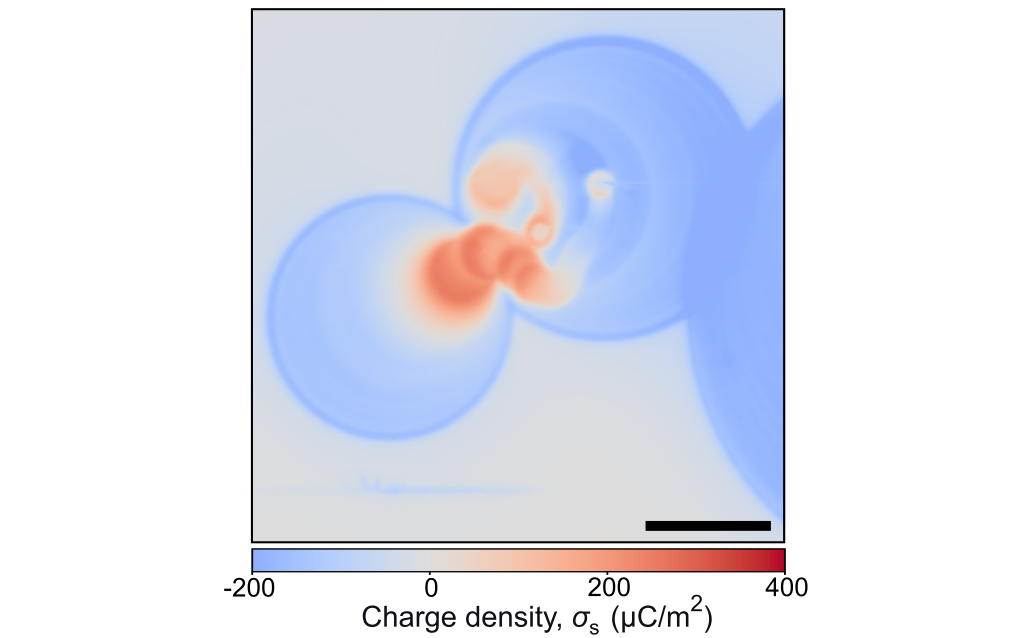}
    \caption{\textbf{Self-propulsion of multiple droplets.} KPFM map of a cluster of neighboring evaporating droplets, showing how each droplet deposits its own distinct charge landscape while also interacting with charge deposited by its neighbors, resulting in more complex, collective charge patterns. Scale bar: \SI{20}{\micro\meter}.
    }
    \label{fig:multiple}
\end{figure}

Fig.~\ref{fig:other_motion} illustrates further examples of self-propelled droplets that remain mostly confined within their initial footprints. For Fig.~\ref{fig:other_motion}(c) and (f), a droplet forms a tight, self-avoiding coil, almost entirely erasing the charge deposited during stage 1, before eventually escaping its initial footprint toward the end of its trajectory. Fig.~\ref{fig:multiple} extends this picture to multiple neighboring droplets: because each droplet leaves behind its own charge landscape, a nearby droplet can be attracted into, and propelled by, charge deposited by its neighbor, producing markedly more complex charge patterns than a single isolated droplet can generate.

\clearpage

%

\end{document}